\documentclass[preprint, twocolumn]{article}

\usepackage[a4paper, margin={1.5cm,1.5cm}]{geometry}

\usepackage{cite}
\usepackage[T1]{fontenc}
\usepackage{amsmath,amssymb,amsfonts}
\usepackage{algorithmic}
\usepackage{graphicx}
\usepackage{textcomp}
\usepackage{listings}
\usepackage{hyperref}
\usepackage{caption}
\usepackage{subcaption}
\usepackage{arydshln}
\usepackage{mathtools}
\usepackage{ragged2e}
\usepackage{array}
\usepackage{multirow}
\usepackage[utf8]{inputenc}
\def\BibTeX{{\rm B\kern-.05em{\sc i\kern-.025em b}\kern-.08em
    T\kern-.1667em\lower.7ex\hbox{E}\kern-.125emX}}
\begin{document}

\title{Semantic Code Graph -- an information model to facilitate software comprehension}

\author{
  Krzysztof Borowski\thanks{AGH University of Krakow and VirtusLab, Virtus Lab Spolka z o.o., Poland (e-mail: kborowski@agh.edu.pl, kborowski@virtuslab.com)} \and 
  Bartosz Balis\thanks{Institute of Computer Science, AGH University of Krakow, Krakow, Poland (e-mail: balis@agh.edu.pl)} \and 
  Tomasz Orzechowski\thanks{R\&D Department, Virtus Lab Spolka z o.o., Krakow, Poland (e-mail: torzechowski@virtuslab.com)}
}




\newcommand{\keywords}[1]{\textbf{Keywords:} #1}

\date{}
\maketitle

\begin{abstract}

Software comprehension can be extremely time-consuming due to the ever-growing size of codebases. Consequently, there is an increasing need to accelerate the code comprehension process to facilitate maintenance and reduce associated costs. A crucial aspect of this process is understanding and preserving the high quality of the code dependency structure. While a variety of code structure models already exist, there is a surprising lack of models that closely represent the source code and focus on software comprehension. As a result, there are no readily available and easy-to-use tools to assist with dependency comprehension, refactoring, and quality monitoring of code. To address this gap, we propose the Semantic Code Graph (SCG), an information model that offers a detailed abstract representation of code dependencies with a close relationship to the source code. We establish the critical properties of the SCG model and demonstrate its implementation for Java and Scala languages. To validate the SCG model's usefulness in software comprehension, we compare it to nine other source code representation models. Additionally, we select 11 well-known and widely-used open-source projects developed in Java and Scala and perform a range of software comprehension activities on them using three different code representation models: the proposed SCG, the Call Graph, and the Class Collaboration Network. We then qualitatively analyze the results to compare the performance of these models in terms of software comprehension capabilities. These activities encompass project structure comprehension, identifying critical project entities, interactive visualization of code dependencies, and uncovering code similarities through software mining. Our findings demonstrate that the SCG enhances software comprehension capabilities compared to the prevailing Class Collaboration Network and Call Graph models. Moreover, the SCG-based data analysis yields actionable software comprehension insights. We also release an open-source tool, \textit{scg-cli}, to assist with result reproduction and further research. We believe that the work described is a step towards the next generation of tools that streamline code dependency comprehension and management.

\end{abstract}




\section{Introduction}
\label{sec:introduction}

Software is becoming increasingly complex and larger, driven by fast-changing business requirements and advances in programming languages, libraries, and runtime infrastructure. At the same time, the quality of software architecture, especially in such a dynamic environment, deteriorates over time \cite{Eick2001}, so programmers need to put growing amount of effort into understanding the source code \cite{Al-Saiyd2017} in order to make required changes or introduce new features important from a product or service perspective.

Multiple measures are applied to facilitate software comprehension by both keeping the code clean and speeding up the code-understanding process. Performing code reviews \cite{McIntosh2016}, following a~written set of conventions \cite{Smit2011}, measuring various software quality metrics \cite{kan2003metrics}, or linting the source code \cite{Prause2008} (e.g. guarding against common anti-patterns) have a positive impact on written code quality. Additionally, Integrated Development Environments (IDEs) help programmers browse, read, comprehend, and maintain code efficiently through common IDE features such as syntax highlighting, code navigation, or advanced refactoring capabilities. Nevertheless, Xia at al. \cite{Xia2018} argue that up to 58 percent of software maintenance activities are related to software comprehension \cite{Siegmund2016, Stapleton2020} and claim that we need more advanced software comprehension tools. With the ever greater importance of software and rapidly growing code bases, even small reduction of time spent on software related activities will bring significant cost savings.  

One of the major code comprehension challenges is understanding and managing dependencies of software entities. \textit{Understanding} code dependencies may comprise finding key code entities\cite{Li2021}, localizing connected elements\cite{Al-Saiyd2017, Alanzi2021}, answering different reachability questions \cite{LaToza2010}, reasoning about entire software structure including software architecture understanding \cite{Savidis2022, Liu2021visual, Falci2017}. \textit{Managing} code dependencies denotes introducing required bug fixes or new features and, at the same time, guarding the project structure against decay signs, such as high coupling between modules or classes \cite{Bois2004}, data abstraction coupling or message passing coupling \cite{Riaz2009}, forming god objects \cite{vaucher2009tracking}, introducing unintended dependencies between code entities, etc. \cite{Bandi2013}. Understanding and managing code dependencies is less time consuming if the source code structure is of high quality. To maintain a~high quality code dependency structure, we can either prevent the erosion process or reorganize the already affected code \cite{Oliveto2011}. Preventing harmful dependencies requires constant code dependency and structure monitoring which, although important from a project cost and maintainability perspective, is not something widely applied. Quality metrics most frequently used in projects, e.g., cyclomatic complexity \cite{McCabe1976} or Lines of Code, focus on a particular method or class and do not asses the quality of code dependency structure. Consequently, proper methods and tools are required to discover, understand, and improve the structure of existing code. 

Regarding research on models for code dependency structure, there exists a body of work demonstrating the utility of software network analysis on various theoretical models \cite{Savidis2022, Li2021, Prieto2020, Alanzi2021, Galindo2020data, Arora2019, Lutellier2018, Porkolab2018, Arora2016, URMA2015, SAVIC2014, Jenkins2007, hyland2006scale, Wheeldon2003, myers2003software}. However, even in the case of widely researched models, such as Call Graph (CG) \cite{Ryder1979, LaToza2010, Kinable2011, Guo2013, Alanzi2021} or Class Collaboration Network (CCN) \cite{Wheeldon2003, Li2021, Bavota2013, Du2021}, the underlying implementations, developed by the researchers to extract and analyze these software models, are rarely published. Consequently, it is very difficult to reproduce the results or pursue further research in the code dependency comprehension field, especially leveraging new or commercial projects. Such a~lack of high quality tools for extracting and analyzing the software dependency structure forces researchers to start from scratch every time. This not only hampers research and efforts, but also makes it impossible to effectively compare different results. Perhaps for this reason we have found no research devoted to comparative empirical evaluation of different code representation models. Furthermore, due to the low emphasis on the practical aspects of previously conducted research, there is little to no impact on software comprehension from the code dependency angle in the commercial software development process. We argue that this is mostly due to the lack of high-quality and well-established stable abstract code dependency model focused on practical usability aspects and, in consequence, tools leveraging such a~model to speed up software comprehension and code refactoring process. Code dependency analysis should not only be confined to a theoretical level but, equally importantly, should be geared towards practical applications, reproduction of results, and further research.

To address the research gap between theoretical code dependency models and their applications, we present \textit{Semantic Code Graph} (SCG), a code dependency information model which captures the structure and semantics of code dependencies in a~software project while preserving the direct relation of the representation to the source code. The \textbf{main objective of our research is to evaluate the effectiveness of the SCG for software comprehension activities}. To this end, we formulate the following two research questions:

\begin{itemize}
    \item \textbf{RQ1}: Does the SCG model enhance software comprehension capabilities in comparison with the Class Collaboration Network and the Call Graph models?
    \item \textbf{RQ2}: Does SCG-based data analysis enable actionable software comprehension insights?
\end{itemize}

To asses the theoretical suitability of the SCG model in software comprehension, we are comparing it to nine other source code representation models. Then, we conduct an empirical study involving eleven popular open source projects. We utilize these models to comprehend project structures, identify critical project entities along with their dependencies, and uncover code similarities. We demonstrate that both CCN and CG can be effectively extracted from SCG model. Furthermore, while each of these models provides a distinct perspective on the project, the SCG stands out as the most comprehensive, enabling analyses that surpass the capabilities of the other two.

The contributions of this work can be summarized as follows:
\begin{itemize}
    \item We define and describe the abstract Semantic Code Graph model and its concrete representations for Java and Scala languages. SCG for Java and Scala is able to describe various types of dependencies between software entities at different levels -- from classes and methods, down to local value definitions and type declarations.
    \item We define and implement SCG intermediate representation and storage format to facilitate code dependency data analysis with external tools and support effective model extraction for new languages.
    \item We conduct an empirical analysis of the SCG model's performance concerning its software comprehension capabilities in comparison to the CCN and CG, and qualitatively analyze its results. 
    \item To enable reproducibility and further research, we publish tools and data used in our research. These include the extracted SCG data for the eleven open-source projects used in the empirical study, and the open-source tool \textit{scg-cli} \footnote{\url{https://github.com/VirtusLab/scg-cli}}. The tool is capable of extracting the SCG data for Java projects. It also supports SCG-based data analysis capabilities, and exporting SCG data for usage in external analysis tools, such as \textit{Jupyter Notebook} or \textit{Gephi}.
\end{itemize}

The remainder of this paper is structured as follows. Section \ref{sec:related-work} presents related work. Section \ref{sec:scg-representation} discusses model requirements and presents SCG formal definition with details for Java and Scala languages. In section \ref{sec:scg-extraction-process}, details of the extraction process and the proposed SCG storage format are presented. In section \ref{sec:scg-comparison} we conduct detailed comparison of SCG model to other established graph software representations. Section \ref{sec:scg-software-comprehension} contains a comprehensive evaluation of the proposed approach by demonstrating diverse software comprehension activities for eleven open-source projects and answers research questions. Section \ref{sec:scg-implications} present practical implications from our empirical study and section \ref{sec:conclusions} concludes the paper.

\section{Related work}
\label{sec:related-work}

In this section we review related work, first focusing on existing code dependency representations and subsequently on tools capable of extracting and analyzing various code dependency models.

\subsection{Code dependency representations}

Early graph representations of programs were focused on program execution and dependencies between code segments. One of the first commonly analyzed code graph models is Control Flow Graph (CFG) \cite{allen1970control} -- a directed graph where nodes are basic program blocks and edges represent control flow paths. CFG allows determining the exact execution order of a~program including different program paths implied by control conditions (like if statements). CFG can be successfully applied to program analysis, such as finding unreachable code, compiler optimizations, code generation, code debugging and others \cite{Nair2020}. CFG focuses on block's execution paths and does not represent dependencies on the statement level. To that end, Ferrante at al. \cite{Ferrante1987} introduced the concept of Program Dependence Graph (PDG) which focuses on single statement dependencies. In PDG code statements are represented as nodes, and edges express two main relations between them:

\begin{itemize}
    \item data dependency (execution of one statement requires data produced by another statement),
    \item control flow (execution of one statement depends on a control condition evaluated by another statement).
\end{itemize}

Control dependencies in PDG are commonly extracted from the Control Flow graph but represent a dependency tree of statements rather than an exact execution path. The PDG was initially used in various code optimizations, such as parallelism detection \cite{Kalyur2016}, code movement, or program slicing \cite{tip1994survey}. This representation can also support program comprehension and maintenance by depicting statement dependency trees in order to assess change impact \cite{acharya2011practical}, find code similarities \cite{Krinke2001, Kim2016, Mehrotra2022, Yu2023}, measure code coverage, or reduce the test set for programs. Another graph-based code model that combines the Control Flow Graph (CFG) and Program Dependence Graph (PDG), along with an Abstract Syntax Tree, is known as the Code Property Graph (CPG) \cite{Yamaguchi2014}. This comprehensive model has been effectively employed in the detection of various vulnerabilities \cite{Backes2017, Suneja2020LearningTM}.

Program Dependence Graph, Control Flow Graph and derivative Code Property Graph, are limited to monolithic programs which means that analysis cannot cross the boundary of procedures. To overcome this drawback, Horwitz et al. \cite{horwitz1988interprocedural} introduced the System Dependence Graph (SDG), augmenting the previous PDG model with edges representing dependencies between a call site and the called procedure, along with handling passed values via "procedure linkages." Tracking values is not trivial to represent as a graph and requires defining several new abstract vertices \cite{Meng2015}. The SDG allows for interprocedural code analysis, such as interprocedural slicing \cite{horwitz1988interprocedural} or test case generation \cite{karuthedath2020}.

Over the years, object-oriented programming and other high-level programming languages have been popularized. Object-oriented programming introduced new concepts such as classes, methods, inheritance, and related abstract dependencies that are not represented in the basic System Dependence Graph. Researchers attempted to address this issue through SDG model extensions, as presented by Walkinshaw et al.~\cite{Walkinshaw2003} in the Java System Dependence Graph (JSysDG). JSysDG is a multigraph that represents both the structure of the program (method headers, classes, interfaces, and packages) and program behavior through SDG. Later, JSysDG was extended with object-flow dependence by Galindo et al.~\cite{Galindo2020data} to better facilitate program slicing. Shu et al.~\cite{Shu_2013} proposed a practical implementation of the JSysDG model in JavaPDG—a platform for program dependence analysis along with a graphical viewer. However, the software solution was not open-sourced, and it is currently impossible to find and use it for pursuing new software dependency research. It is also worth mentioning that JSysDG builds on a similar concept to the Java Software Dependence Graph researched by Zhao~\cite{Zhao2001}.

Arora et al.~\cite{Arora2012} summarize various variations of Control Flow Graphs and Program Dependence Graphs. These models primarily aim to represent program execution flow or statement dependencies. The structure of the program, including abstract code elements like classes, is added on top of these graphs to facilitate more detailed program execution analysis and adapt to higher-level programming languages. This implies that software is analyzed at a very detailed level of statements, with a focus on understanding and optimizing program execution. However, for programmer, this level of detail may not be as critical with modern languages. Many sophisticated optimizations are handled directly by compilers or interpreters, reducing the programmer's responsibility. Moreover, modern languages operate at higher abstraction levels, introducing abstract components and focusing on language expressiveness, better ways of structuring growing software, enabling component reuse, facilitating implementation replacements, and simplifying local refactoring activities. This shift in focus towards higher abstraction levels allows for faster program development, reduces the occurrence of certain errors, and enhances maintainability. Changing requirements, increasing software criticality, and the significant growth in codebase sizes have driven research efforts towards software structure comprehension and maintainability. The aim is to reduce the time and cost associated with software development.

The Call Graph \cite{Ryder1979} is another popular approach wherein the graph of caller-callee relationships is extracted from the source code. The main goal of the Call Graph is to support static and dynamic analysis of the code call dependency flow without concerning the detailed code structure. Grove et al. \cite{Grove1997} distinguish two types of the Call Graph -- the context-insensitive one, in which one node uniquely represents one procedure, and the context-sensitive one, in which each procedure call is represented as a~separate node. For code navigation and depicting the graph structure, the context-insensitive approach is more natural, as each definition in the code has precisely one unique node representation, similar to the source code definition in the file. Various papers use call graphs for code analysis \cite{LaToza2010, Kinable2011, Guo2013, Alanzi2021}. Although proven useful, the Call Graph is not a~complete representation of the code dependencies as it contains only information about chains of calls.

Software networks, also referred to as Artifact Dependency Graphs (ADG) \cite{Isazadeh2017}, software collaboration graphs \cite{myers2003software}, or software architecture graphs \cite{Jenkins2007}, are general terms for abstract source code models that focus on code entities and their dependencies. They are not intended to depict program execution but rather to describe the abstract graph model of the program's structure and dependencies. Myers \cite{myers2003software} demonstrates that such graphs form scale-free, small-world networks and can be analyzed using a set of algorithms and metrics known from other complex network analyses \cite{Jenkins2007, Gomez2019}.

For object-oriented programming languages, software networks can consist of multiple levels of networks, such as file-level, package-level, class-level, or function-level collaboration networks \cite{SAVIC2014}. The Class Collaboration Network (CCN) is particularly interesting to analyze \cite{Wheeldon2003, Li2021, Bavota2013, Du2021} because classes are the main building blocks in OOP. Arora and Goel \cite{Arora2019, Arora2016} proposed JavaRelationshipGraphs (JRG) — a software network model specialized for the Java language. JRG is a directed graph that captures various relations between different code elements in Java programs. In this graph, nodes represent packages, classes, interfaces, methods, attributes, and edges represent class extensions, implementations and aggregations, imports, ownership, and method calls. There is not one definitive guide on how to create such a software network for a particular language, but there is an attempt made by Savic et al. \cite{SAVIC2014} to create a language-independent model of dependencies between source code entities and establish a General Dependency Network (GDN). While this is a promising direction, it may miss some language-specific details and GDN is extracted from yet another custom model — the extended abstract syntax tree (eCST), which complicates model adoption and potential implementations.

Software networks are extensively used to facilitate software comprehension and maintenance through software architecture recovery \cite{Savidis2022, Lutellier2018}, finding key entities \cite{Meyer2015, Pan2018, Li2021, Du2021}, code modularization \cite{Srinuvasu2016, Isazadeh2017, Pourasghar2021}, finding code similarities \cite{Tyagi2022}, code visualization \cite{Mattila2016, Porkolab2018codecompass, Borowski2022}, and tracking code evolution \cite{Bhattacharya2012}.

\subsection{Tools for extracting and analyzing code dependencies}

To leverage the progress made in software comprehension research related to program graph representation, the utilization of readily available code dependency graph representation tools is essential. Furthermore, to facilitate ongoing research and establish persuasive comparisons, the importance of publicly accessible graph extraction solutions is crucial. These solutions should export graph models in an easily digestible intermediate data representation. It is rather surprising that, even for widely used languages such as Java, a shortage of such tools is evident. For the Java language, \textit{Scoot} \cite{vallee2010soot, lam2011soot}, and its successor \textit{SootUp}\footnote{\url{https://soot-oss.github.io/SootUp/}}, although primarily created for introducing bytecode optimizations, can generate Control Flow Graphs (for particular methods or classes) and extract Call Graphs for the project through the Java API.

\textit{JDepend}\footnote{\url{https://github.com/clarkware/jdepend}} can generate predefined design quality metrics, but it does not expose an internal model that could be used for other software comprehension activities.

\textit{Dependency Finder}\footnote{\url{https://depfind.sourceforge.io/}} can find and export Java simplified dependencies from bytecode representation, but it can only extract three types of dependencies: class-to-class, feature-to-class, and class-to-feature (where feature represents everything but a class). \textit{Dependency Finder} can export dependencies to TXT or XML format. However, aside from the fully qualified name of the entity, no other details are stored in this model.

\textit{jdeps}\footnote{\url{https://docs.oracle.com/en/java/javase/11/tools/jdeps.html}} is another tool capable of presenting Java dependencies extracted from bytecode. However, these dependencies are limited to only package and class levels and do not consider other code entities or precise relations between them.

\textit{Scitools Understand}\footnote{\url{https://scitools.com/}} allows for various static code analyses, code dependency visualization, and advanced code browsing. However, \textit{Scitools Understand} is a paid tool for commercial usage and can only export basic class-level and file-level software networks.

Code Compass \cite{Porkolab2018codecompass} is an open-source LLVM/Clang-based tool developed to augment the understanding of large legacy systems. It supports call graph and class collaboration networks visualization and has some browsing capabilities but does not support any software network export capabilities.

Wheeldon and Counsell \cite{Wheeldon2003} in their work extract a detailed class collaboration network with \textit{AutoCode} software, but its actual implementation is not available.

Based on our current knowledge, there is no publicly available and free to use tool to extract a rich System Dependence Graph or Software Network Graph for the Java or Scala languages. 

Additionally we conduct a detailed comparison of existing model with SCG in section \ref{sec:scg-comparison}. We also point out in details shortcomings of existing models from the software comprehension perspective. To the best of our knowledge, there has been no comparative study focusing on graph code representations in the context of software comprehension.

\section{Semantic Code Graph}
\label{sec:scg-representation}

\subsection{Source code dependency model requirements}
\label{sec:scg-requirements}

The goal of the Semantic Code Graph information model is to preserve the direct relation to the source code (syntax) while also capturing the meaning (semantics) of the source code elements and dependencies between them.  An information model that meets such requirements will facilitate source code analysis in various areas. Here we particularly focus on three of them:

\begin{itemize}
    \item \textit{Software comprehension}, e.g., learning code structure by interactive code visualization and browsing, applying data analysis to find the most important code elements, answering reachability questions \cite{LaToza2010}, or facilitating semantic code search \cite{Husain2019}.
    
    \item \textit{Software quality assessment}, e.g., measuring software metrics through graph properties \cite{Falci2017, Jenkins2007, Rosenberg1995, Prause2008}.
    
    \item \textit{Software refactoring}, e.g., supporting software modularization through suggested code partitioning, suggesting method movements, or code structure improvements \cite{Bois2004, Oliveto2011, Brito2020}.
\end{itemize}

The important `direct relation to source code' requirement simply denotes that the entities seen by the developer while working with the source code are represented in the model as closely as possible. Consequently, the SCG model should be extracted either directly from the source code, or another representation that closely matches the source code, e.g. the Abstract Syntax Tree (AST). Extraction from an intermediate representation, such as bytecode -- while tempting for various reasons -- would not be appropriate since such a representation contains entities generated by the compiler (see section \ref{sec:bytecode} for more in-depth ana\-lysis of this). Moreover, all entities and relations between entities should have the \textit{location} property associated with them. This is a crucial property for building interactive visualization, or for pointing the developer to the exact place in the source code when performing static analysis or preparing any potential code manipulations. Any advice presented by a tool -- like presenting visual structure, found problems, extracted knowledge, or improvement suggestions -- should always precisely point the user to an appropriate place in the source code.

In addition, each extracted entity and relation should retain as much semantic information as possible. For example, an entity representing a~method should also contain its scope (private, public), modifiers (abstract, override, etc.), the number of arguments and their types, return types, etc. This will allow for precise, close-to-source-code model analysis focused on dependencies between source code entities, at the same time leveraging most of details available in the AST code representation.


\subsection{Semantic Code Graph base model}
\label{semantic-code-graph-definition}

Semantic Code Graph (SCG) is a~source code information model aimed at representing diverse dependencies present in the source code, with the objective of fulfilling the requirements outlined in Section \ref{sec:scg-requirements}. The SCG data structure is a~graph where nodes denote code declarations and definitions, while the edges represent various dependencies between them. The SCG is formally defined as a pair $G_{SCG}=(V_D, E_R)$ comprising:
\begin{itemize}
    \item $V_{D} = \{ v_1, v_2, \ldots, v_n \}$, a set of vertices (nodes), where $v_i$ represents a code entity of type \textit{CLASS}, \textit{OBJECT}, \textit{TRAIT}, \textit{INTERFACE}, \textit{METHOD}, \textit{PARAM}, \textit{TYPE\_PARAM}, \textit{VALUE}, \textit{VARIABLE}, \textit{TYPE}, \textit{ENUM}. This list of node types is flexible and can be extended with other entities specific for given language; $n$ is the total number of entities;
    \item $E_R = \{ (v_i, v_j) \mid v_i, v_j \in V_D \}$, a set of directed edges comprising ordered pairs of nodes, and representing \textit{relations} between these nodes. The possible types of edges include \textit{CALL}, \textit{DECLARATION}, \textit{EXTEND}, \textit{OVERRIDE}, \textit{PARAMETER}, \textit{RETURN\_TYPE}, \textit{TYPE\_PARAMETER}. Similar to nodes, new edge types can be added if required to capture specific code dependency relations occurring in a~language.
\end{itemize}
The basic data structures of the SCG graph, as a~foundation for diverse semantic information about source code dependencies, are rather simple. In particular, each node has to contain the following:
\begin{itemize}
    \item \textit{id} -- unique node identifier,
    \item \textit{kind} -- type of the node (one of the node types mentioned in the $V_D$ definition above),
    \item \textit{location} -- place in the source code where the entity represented by this node is defined, 
    \item \textit{displayName} -- node name meaningful to the end user,
    \item \textit{edges} -- set of outgoing edges.
\end{itemize}
The edges in the SCG graph are described by the following properties:
\begin{itemize}
    \item \textit{to} -- node identifier to which this edge is pointing to,
    \item \textit{location} -- place in the source code where the relation represented by this edge can be observed,
    \item \textit{type} -- the type of the edge (its label); \textit{type} is one of the edge types mentioned in the $E_R$ definition above.
\end{itemize}

The \textit{location} property has its own structure that provides precise information about the position of a node or edge in the source code file. It comprises the following fields:
\begin{itemize}
    \item \textit{uri} -- the relative path to the source code file,
    \item \textit{startLine} -- indicates the starting line number for the position of the node or edge in the file,
    \item \textit{startCharacter} -- specifies the starting character position for that node or edge,
    \item \textit{endLine} -- indicates the end line number for the position of that node or edge in the file,
    \item \textit{endCharacter} -- specifies the ending character position for that node or edge.
\end{itemize}

The SCG nodes represent all concrete entities that describe the essential code structure. In particular, nodes do not represent abstract or arbitrary code organization entities, such as packages, files, or modules. The information about these aspects (e.g. that a class belongs to a package or is defined in a certain file) is very useful, but it is better to provide it as additional node or edge properties, rather than essential code structural elements, as it would disrupt the representation of the software structure and quality analysis of the code.

In addition to essential properties, each node and edge can contain various additional properties encapsulated in the \textit{properties} map field. These properties may encompass information such as the number of lines of code (LOC), the package to which the entity belongs, its visibility scope, various modifiers, or any language-specific details that prove useful for subsequent analysis. Furthermore, properties can be extracted from sources other than the source code itself, for example, from the source control version system, to acquire details such as the last edit timestamp, the primary author, the latest commit message, and so forth. However, it should be noted that mapping certain source control version information into SCG nodes is not straightforward because most source control versions operate at the level of entire lines. When dealing with classes, extracting the author of the most recent changes becomes feasible by utilizing the class declaration location and the property indicating the number of lines of code within the class. This allows us to search for the last change within the class's defined scope. For one-line definitions such as variables or parameters, we can take advantage of the token-based git history approach introduced by the Context Buddy tool\footnote{\url{https://github.com/VirtusLab/contextbuddy/tree/master/intellij}}. This tool explores the git history from the project's inception and, among other things, extracts and tracks the introduced tokens, token history, and the original authors of the tokens. Token history can be easily matched with a specific node, thanks to the node's \textit{location} property.

The specific node kinds and edge types, along with their connections, are language-specific details that should be implemented in a manner natural and most suitable for each programming language. In the following sections, we describe specific implementations of the SCG model for the Scala and Java languages.

\subsection{Semantic Code Graph for Scala and Java}

Scala \cite{Odersky2004} is a hybrid language that combines constructs known from object-oriented and functional programming. Scala 2 and its next iteration Scala 3, themselves based on JVM, were designed to supersede the Java language in its capabilities, extending and improving on already proven useful features rather than completely changing the paradigm. As a result, Java and Scala share similar entity types and relations in terms of the Semantic Code Graph, as presented in Table \ref{fig:scg-java-scala-types}. 

\begin{table*}[!h]
\begin{footnotesize}

\begin{tabular}{p{0.7cm}|p{3cm}|>{\raggedright\arraybackslash}p{6cm}|>{\raggedright\arraybackslash}p{6cm}}

    Java/ Scala & Node kind & Description & Example \\
    \cline{1-4}
    J,S &\textit{CLASS} & a class & \texttt{class A} \\
    S   &\textit{OBJECT} & an object & \texttt{object A} \\
    S   &\textit{TRAIT} & a trait &\texttt{trait B} \\
    J   &\textit{INTERFACE} & an interface & \texttt{interface B}  \\
    J,S &\textit{METHOD} & a method &\texttt{def m(): A}  \\
    J,S &\textit{PARAM} & method, constructor or trait parameter & \texttt{a} in \texttt{def m(a: A)} \\
    J,S & \textit{TYPE\_PARAM} & a type parameter & \texttt{T} in \texttt{def m[T](t: T)} \\
    J,S &\textit{VALUE} & value in Scala or final variable in Java & \texttt{val a: A} \\
    J,S &\textit{VARIABLE} & a variable (non-final field) &\texttt{var a: A} \\
    S   &\textit{TYPE} & type member & \texttt{type T $<$: A} \\
    J,S &\textit{ENUM} & enumerated type & \texttt{enum E \{ case A \}} \\
    \cline{1-4}
    Java/ Scala & Edge type & Description &  Example\\
    \cline{1-4}
    J,S & CALL & call to method, constructor; usage of parameter or field & \texttt{def m1()=m2()} $\Rightarrow$  $m1\xrightarrow{CALL}m2$\\
    J,S & DECLARATION & relation between parent and declared entity & \texttt{class A \{ def m()=() \}} $\Rightarrow$  $A\xrightarrow{DECLARATION}m$ \\
    J,S & EXTEND & extend relation between entities & \texttt{class A extends B} $\Rightarrow$ $A\xrightarrow{EXTEND}B$ \\
    J,S & OVERRIDE & edge from (implementation) method to its overridden method & \texttt{override def m()} $\Rightarrow$ $A\#m\xrightarrow{OVERRIDE}B\#m$ \\
    J,S & PARAMETER & edge from entity to its parameter & \texttt{def m(a: A)} $\Rightarrow$ $m\xrightarrow{PARAMETER}a$ \\
    J,S & RETURN\_TYPE & edge from method to its return type & \texttt{def m(): B} $\Rightarrow$ $m\xrightarrow{RETURN\_TYPE}B$ \\
    J,S & TYPE & type relation e.g. between variable definition and its type & \texttt{val a: A} $\Rightarrow$ $a\xrightarrow{TYPE}A$ \\
    J,S & TYPE\_PARA\-METER & type parameter & \texttt{T} in \texttt{def m[T](t: T)} $\Rightarrow$  $m\xrightarrow{TYPE\_PARAMETER}T$ \\

\end{tabular}
\end{footnotesize}
\caption{List of nodes and edges present in SCG model for Java and Scala}
\label{fig:scg-java-scala-types}
\end{table*}

\begin{figure}[!h]
\centering
\begin{subfigure}[b]{0.4\textwidth}
  \centering
  \begin{lstlisting}[basicstyle=\footnotesize]
trait A {
  def foo(): Unit = ()
}
class B extends A {
  def bar(): Unit = foo()
}
  \end{lstlisting}
  \caption{Scala simple inheritance example.}
\end{subfigure}%

\begin{subfigure}[b]{0.4\textwidth}
  \centering
  \includegraphics[width=200px]{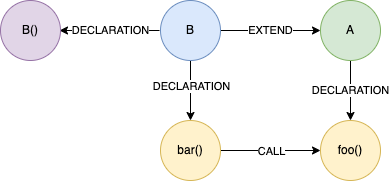}
  \caption{Semantic Code Graph visualized.}
\end{subfigure}
\caption{SCG visualization for a simple inheritance example.}
\label{fig:scala-trait-inheritance-scg}
\end{figure}

\begin{figure}[!htb]
\centering
\begin{subfigure}[b]{0.4\textwidth}
  \centering
  \begin{lstlisting}[basicstyle=\footnotesize]
1. def triple(n: Int): Int = {
2.  val t = n * n * n
3.  t
4. } 
  \end{lstlisting}
  \caption{Simple method with local value and method parameter called multiple times.}
\end{subfigure}%
\hfill
\begin{subfigure}[b]{0.4\textwidth}
  \centering
  \includegraphics[width=200px]{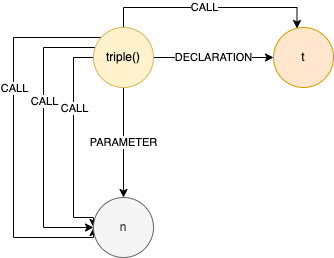}
  \caption{Semantic Code Graph visualized.}
\end{subfigure}
\caption{SCG visualization of method with local calls and local value \texttt{t} definition.}
\label{fig:scala-local-variable-scg}
\end{figure}

Fig. \ref{fig:scala-trait-inheritance-scg} and Fig. \ref{fig:scala-local-variable-scg} show source code samples and their respective SCG representations. Fig. \ref{fig:scala-trait-inheritance-scg} presents two top-level declarations of class \textit{B} and trait \textit{A}. The respective SCG graph captures various dependencies in this piece of code depicting each declared entity -- class \textit{B} with method \textit{bar()}, trait \textit{B} with method \textit{foo()}, and existing dependencies between them of type \textit{DECLARATION}, \textit{EXTEND} and \textit{CALL}.

Fig. \ref{fig:scala-local-variable-scg} shows method \textit{triple} with one parameter \texttt{n} and local value \texttt{t} which is computed as \texttt{$n*n*n$}. Consequently, the \textit{n} parameter of method \texttt{triple} is called by this method 3 times. The \texttt{t} parameter is then called and returned in line 3. Type \texttt{Int} is deliberately not represented in the SCG as its definition exists only in an external library. SCG represents only entities defined within the code base of the analyzed project.

\section{Semantic Code Graph extraction process}
\label{sec:scg-extraction-process}

\subsection{Implementation Overview}
\label{scg-implementation-overview}

A common way of extracting the structure of the source code is by parsing it into an Abstract Syntax Tree (AST). However, the AST represents only syntax, lacking the semantic information about relations between code entities, required by the SCG (\ref{semantic-code-graph-definition}). Also, the AST describes entities not important from the semantic code dependency point of view, such as braces, keywords, and comments. To build a~comprehensive SCG model from an AST, we therefore need to drop unnecessary information from the AST graph and augment the remaining nodes with semantic information. Adding relations between entities will create a graph of semantic and structural dependencies. 

\begin{figure}[!htb]
    \centering
    \includegraphics[width=200px]{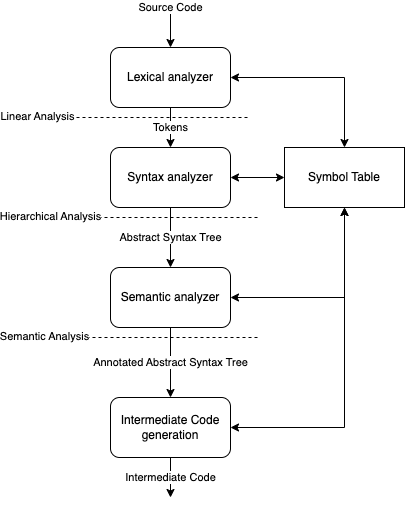}
    \caption{Front-end analysis steps in a compiler}
    \label{fig:compiler-fe}
\end{figure}

Semantic analysis is one of the crucial steps in the compilation process \cite{Grune2012-intro}. Consequently, useful semantic information required by the Semantic Code Graph can be extracted from the information generated during the semantic analysis performed by the compiler in the front-end analysis phase of the compilation. All steps of the front-end analysis are schematically depicted in Fig. \ref{fig:compiler-fe}. In the \textit{Hierarchical Analysis} step, the parser consumes code tokens and produces an Abstract Syntax Tree, a lossless representation of the source code, preserving all its syntactical details. In the next step, the semantic analyzer augments the AST with additional semantic and contextual information with the help of \textit{Symbol Tables}. Symbol Tables facilitate efficient insertion and lookup of code elements by name, reflecting code dependencies, e.g., linking references to their corresponding declarations. \textit{Semantic Analysis} (also known as context-sensitive analysis) annotates the AST with types, references, scopes, and other semantic attributes. Such an AST, with all types and symbols resolved, is a sufficiently detailed abstraction for generating SCG.

\subsection{Extracting SCG for the Scala2 language}
\label{scg-scala2}

The Scala 2 compilation process consists of 24 phases (Fig. \ref{fig:scalac-compiler-phases}), of which only the first four (\textit{parse}, \textit{namer}, \textit{packeobjects} and \textit{typer}) are required to extract information useful for the SCG. To this end, we can create a~compiler plugin and work directly with the compiler internal structures. However, compilation is a complex process, so dealing with compiler internal structures can be challenging. An alternative solution is to use other meta-tools implemented to facilitate Scala 2 syntactic and semantic analysis. An untyped AST can be generated with the scalameta \textit{Trees}\footnote{\url{https://scalameta.org/docs/trees/guide.html}} library and then annotated with semantic information available through the \textit{SemanticDB}\footnote{\url{https://scalameta.org/docs/semanticdb/guide.html}} library, which exposes symbols and types defined in the source code. \textit{SemanticDB} is a compiler plugin that is attached after the typer phase. We can use these two libraries and create our own Scala plugin\footnote{\url{https://github.com/VirtusLab/scg-scala}} capable of extracting SCG, also invoked after the typer compilation phase. 

\begin{figure}[!htb]
  \centering
  \footnotesize
  \begin{verbatim}
$ scalac -Xshow-phases
    phase name  id  description
    ----------  --  -----------
        parser   1  parse source into ASTs, 
                    perform simple desugaring
         namer   2  resolve names, 
                    attach symbols to named trees
packageobjects   3  load package objects
         typer   4  the meat and potatoes: 
                    type the trees
         ...
           jvm  23  generate JVM bytecode
      terminal  24  the last phase 
                    during a compilation run
  \end{verbatim}
  \caption{Scala 2 compiler phases. All type related information are available after the \textit{typer} phase.}
  \label{fig:scalac-compiler-phases}
\end{figure}

\subsection{Extracting SCG for the Scala3 language}
\label{scg-scala3}

The Scala 3 compiler introduces the Typed Abstract Syntax Tree (TASTy) \cite{Odersky2016-tasty-reference}, an annotated AST serving as an intermediate code between the front and back end of the compiler. TASTy contains a~complete source code syntax and resolved semantic details (such as types, references, and others). The TASTy trees are stored as binary files \textit{*.tasty} during the \textit{pickler} compilation phase (Fig. \ref{fig:scala3c-compiler-phases}). Instead of writing our own compiler plugin, we can read these files from the compiled project and generate the SCG in an external process. 

\begin{figure}[!htb]
  \centering
   \footnotesize
  \begin{verbatim}
phase name  description
----------  -----------
    parser  scan and parse sources
     typer  type the trees
       ...
   pickler  generates TASTy info
       ...
  genBCode  generate JVM bytecode
    \end{verbatim}
    \caption{Scala 3 compiler phases, with \textit{pickler} phase for storing TASTy binary files.}
  \label{fig:scala3c-compiler-phases}
\end{figure}

\subsection{Extracting SCG for the Java language}
\label{scg-java}

The extension mechanism for the \textit{javac} compiler was introduced in Java 8. Over the years, writing compiler plugins and dealing with compiler internals has proven to be challenging and subject to internal API changes. To facilitate code analysis that does not require full features of the compiler, the \textit{JavaParser}\footnote{\url{http://javaparser.org/}} library was created. \textit{JavaParser} not only produces the AST, but also (since version 3.5.0) resolves semantic dependencies through a~special module called \textit{SymbolSolver}. This feature can be utilized to extract the SCG from the Java source code. The library supports all language constructs from Java 1 to Java 15 and is actively developed, which gives us great coverage of language features with minimal effort. We have enhanced our SCG command line tool, \textit{scg-cli} (introduced in Section \ref{subsec:scg-cli-intro}), with the functionality to extract SCG data for Java projects. This feature can be accessed using the command \texttt{scg-cli generate <path/to/project>}.



\subsection{Challenges in extracting SCG from Bytecode}
\label{sec:bytecode}

Sections \ref{scg-scala2}-\ref{scg-java} described the implementation of the SCG extraction for three different languages: Scala 2, Scala 3 and Java. All three of them, with the help of compilers during the back-end analysis, are translated to the bytecode -- the Intermediate Code representation for the Java Virtual Machine. Consequently, it would be beneficial from the implementation point of view to extract SCG from the JVM bytecode as this would automatically support any JVM-based language. In section \ref{scg-implementation-overview}, we discussed how the compiler front-end provides the most convenient and accurate representation of the source code for our objectives. On the contrary, bytecode is an output from the compiler back-end process which involves various optimizations and simplifications applied to generate an output optimized for the JVM interpreter. Indeed, the accuracy of the model extracted from bytecode suffers from representation mismatch. Let us consider a~few notable examples:
\begin{itemize}
    \item Missing information on precise element location in the source code.
    \item Missing type information due to the type erasure mechanism.
    \item Additional JVM bytecode generated to cover specific language features not represented in bytecode.
    \item Various compiler optimizations.
\end{itemize}

\subsubsection{Missing location data}
Java programs are compiled with the \textit{javac} compiler. The resulting bytecode contains only limited information on the location of the software entities. Even with the \textit{-g} parameter, the generated bytecode does not contain location for non-executable code such as abstract methods (Fig. \ref{fig:java-interface-missing-location}).

\begin{figure}[!htb]
\centering
\begin{subfigure}[b]{0.45\textwidth}
  \centering
  \begin{lstlisting}[basicstyle=\footnotesize]
//A.java
1. public interface A {
2.     void f1();
3. }
4.
5. class B implements A {
6.   public void f1() {
7.  }
8. }
  \end{lstlisting}
  \caption{Java \textit{A} interface with \textit{f1} method and its class \textit{B} implementation.}
\end{subfigure}%
\hfill
\begin{subfigure}[b]{0.45\textwidth}
  \centering
  \begin{lstlisting}[basicstyle=\footnotesize]
$ javac -g A.java  
$ javap -l A 
Compiled from "A.java"
public interface A {
  public abstract void f1();
}
$ javap -l B
Compiled from "A.java"
class B implements A {
  B();
    LineNumberTable:
      line 5: 0
    ...

  public void f1();
    LineNumberTable:
      line 7: 0
    ...
  \end{lstlisting}
  \caption{Decompiled bytecode -- missing location for the interface and its methods; line number only generated for constructor \texttt{B} and method \texttt{f1()}.}
\end{subfigure}
\caption{Java bytecode does not contain the source code location of non-runtime code elements even with the \texttt{javac -g} flag. For other elements only the line number can be recorded.}
\label{fig:java-interface-missing-location}
\end{figure}

\subsubsection{Missing type data}
In Java, we can define generic types, such as \textit{TypeErasure$<$A$>$}, but the actual type will be lost in the compilation due to type erasure mechanism\footnote{\url{https://docs.oracle.com/javase/tutorial/java/generics/erasure.html}} (Fig.\ref{fig:type-erasure}).  

\begin{figure}[!htb]
\centering
\begin{subfigure}[b]{0.45\textwidth}
  \centering
  \begin{lstlisting}[basicstyle=\footnotesize]
//TypeErasure.java
class TypeErasure<A> {
    public A a;
    public TypeErasure(A a){
        this.a = a;
    }
}
//M.java
public class M {
    void main(String[] args) {
        TypeErasure<Integer> a = 
            new TypeErasure<Integer>(1);
        System.out.println(a.a + 1);
    }
}
  \end{lstlisting}
  \caption{\textit{TypeErasure$<$A$>$} generic type}
\end{subfigure}%
\hfill
\begin{subfigure}[b]{0.45\textwidth}
  \centering
  \begin{lstlisting}[basicstyle=\footnotesize]
//TypeErasure.class
class TypeErasure<A> {
    public A a;

    public TypeErasure(A var1) {
        this.a = var1;
    }
}
//M.class
public class M {
    public M() {
    }

    public void main(String[] var0) {
        TypeErasure var1 = new TypeErasure(1);
        System.out.println(
            (Integer)var1.a + 1
        );
    }
}
  \end{lstlisting}
  \caption{Decompiled bytecode}
\end{subfigure}
\caption{Java source code: The \textit{Integer} type for the generic \textit{TypeErasure} class will not be represented in bytecode for the local variable \textit{a}.}
\label{fig:type-erasure}
\end{figure}

\subsubsection{Extra compiler-generated bytecode}
The problem of inadequate JVM bytecode is especially visible for languages other than Java. For example, Scala has a~\textit{trait}\footnote{\url{https://docs.scala-lang.org/tour/traits.html}} code structure that is similar to a Java interface, but it can have concrete fields and methods, and provides the multiple inheritance capability with the help of the Scala trait linearization algorithm (Fig. \ref{fig:scala-trait-inheritance-java-bytecode}). 

\begin{figure}[!htb]
\centering
\begin{subfigure}[b]{0.45\textwidth}
  \centering
  \begin{lstlisting}[basicstyle=\footnotesize]
trait A {
  def foo(): Unit = ()
}

class B extends A {
  def bar(): Unit = foo()
}
  \end{lstlisting}
  \caption{Scala source code: class B extends trait A.}
\end{subfigure}%
\hfill
\begin{subfigure}[b]{0.45\textwidth}
  \centering
  \begin{lstlisting}[basicstyle=\footnotesize]
// A.class
public interface A {
   static void foo$(final A a) {a.foo();}
   default void foo() {}
   static void $init$(final A a) {}
}

// B.class
public class B implements A {
   public void foo() { A.foo$(this); }
   public void bar() {this.foo();}
   public B() {A.$init$(this);}
}
  \end{lstlisting}
  \caption{Decompiled compiler-generated bytecode.}
\end{subfigure}
\caption{The Scala compiler generates the \textit{B\#foo()} method with reference to the generated static method \textit{A\#foo\$(a)}.}
\label{fig:scala-trait-inheritance-java-bytecode}
\end{figure}

\begin{figure}[!htb]
    \centering
    \includegraphics[width=220px]{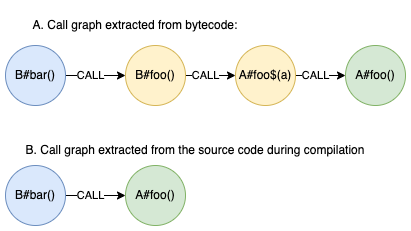}
    \caption{Comparison of a call graph generated from bytecode and from the source code -- \textit{B\#foo()} and \textit{A\#foo\$(a)} methods do not exist in the source code.}
    \label{fig:call-graph}
\end{figure}

Even in very simple cases, the call graph generated from the Scala bytecode can be very different from one that actually occurs in the source code, as presented in Fig. \ref{fig:call-graph}.

Such extensive changes to the generated bytecode result in a mismatch between the source code and its representation in the bytecode. 
Consequently, accurate information about code dependencies can only be extracted directly from the source code rather than using an intermediate code representation such as bytecode.

\subsection{Stable symbol identifiers for a multi-language SCG model}
\label{scg-stable-system-identifiers}

To create bindings between references and definitions, and reflect them in a graph structure, we need to adopt conventions for constructing global stable identifiers for code entities. The same identifier has to be generated when analyzing a~definition and when analyzing a~reference to this definition. The unique symbol identifier can also be used as a node identifier and to construct appropriate relations between nodes in the SCG. 
Symbol names are unique only within a~given context, and some names might be shadowed in an inner context. To create unique identifiers, we need to combine the symbol name with that symbol's owner identifier. In this way, we can ensure the uniqueness of each symbol identifier. We propose a~set of rules\footnote{partially adopted from \url{https://scalameta.org/docs/semanticdb/specification.html\#symbolinformation-1}. Access 22.05.2023.} presented in Table \ref{fig:construction-rules}, using which we can construct a stable identifier for any language symbol.



\begin{table}[!htb]
\centering
\begin{small}
\setlength{\tabcolsep}{3pt}
\begin{tabular}{r|r|l|l}

    Lang & Symbol type & Rule & ID \\
    \cline{1-4}
    J,S & \textit{PACKAGE} & name + \textit{/} & \textit{p/} \\
    J,S &\textit{CLASS} & Owner ID + \textit{\#} & \textit{p/A\#} \\
    S &\textit{OBJECT} & O. ID + \textit{.} & \textit{p/B.} \\
    S &\textit{TRAIT} & O. ID + \textit{\#} & \textit{p/T\#} \\
    J,S &\textit{METHOD} & O. ID + name + \textit{().} & \textit{p/B.mB().} \\
    J,S &\textit{METHOD*} & O. ID + name + \textit{(+n).} & \textit{p/B.mB(+1).} \\
    J,S &\textit{PARAM} & O. ID + \textit{(name)} & \textit{p/A\#mA().(a)} \\
    J,S & \textit{TYPE\_PARAM} & O. ID+ \textit{[name]} & \textit{p/A\#mT().[T2]} \\
    J,S &\textit{VALUE} & O. ID + name + \textit{.} & \textit{p/B.b.} \\
    J,S &\textit{VARIABLE} & O. ID + name + \textit{().} & \textit{p/B.c().} \\
    S &\textit{TYPE} & O. ID + name + \textit{\#} & \textit{p/B.T\#}

\end{tabular}
\end{small}
\caption{Rules for creating stable identifiers in the Java and Scala languages (related code containing all symbol types is presented in Listing \ref{lst:construction-rules}). *Note that overloaded methods are tagged with a~position number.}
\label{fig:construction-rules}
\end{table}

\begin{figure}[!htb]
 \begin{lstlisting}[basicstyle=\footnotesize, language=Scala, caption=Scala code reflecting stable identifiers presented in Table \ref{fig:construction-rules}, label={lst:construction-rules}]
// A.java
package p
class A {
  def mA(a: String): String = a 
  def mT[T2](t: T2): Unit = {}
}

// B.scala
package p
object B {
  type T = String
  val b = "b"
  var c = "c"
  def mB(a: A): String = {
    a.mA(b) 
  }
  def mB(a: String): String = {
    ""
  }
}
\end{lstlisting}
\end{figure}

\subsection{SCG storage format}
\label{sec:SCG-storage-format}

It is essential to store the SCG data in a format optimized for performance, both during data generation and analysis. We have identified key attributes that such a format should support: 
\begin{enumerate}
    \item \emph{Incremental builds} -- some software projects may contain millions lines of code, therefore the format has to allow building the graph incrementally during the generation process \cite{cserep2020integration}.
    \item \emph{Partial graph loading} -- exported graphs may contain hundreds of thousands of vertices, so it would be inefficient to load the entire graph in the case the user needs to perform analysis only on a~particular module.
    \item \emph{Minimum disc space usage} -- disk usage is important as it will be desirable to distribute graph files in the control version system or as an attachment to the particular code base.
    \item \emph{Extensibility} -- To support various programming languages, we need to be able to add different properties to the graph, so the format needs to be easily expandable.
    \item \emph{Libraries in multiple programming languages} -- Code generating the SCG data for a given language will typically be written in the same language. On the other hand, the code to analyze the SCG data may be written in different suitable languages, such as Python or R. Consequently, it is crucial to provide APIs to write and read the format in multiple programming languages.
    \item \emph{Conversion to other data formats} -- for more advanced data analysis we need to be able to convert the SCG representation to other data formats, such as data frames, or graph file formats specific for existing tools, e.g., as GDF\footnote{\url{https://www.iso.org/standard/54610.html}} or DOT\footnote{\url{https://graphviz.org/doc/info/lang.html}}). 
\end{enumerate}

Many different graph formats were proposed for exporting huge graph structures \cite{Yousfi2014}, but none of them satisfies all our requirements. Consequently, we propose a new graph format based on the Protocol Buffers\footnote{\url{https://developers.google.com/protocol-buffers}}. For the Semantic Code Graph this format can be treated as an intermediate data exchange format between the source code and a~data analysis tool, or a~target graph format file. To address the above requirements, we made the following design choices:
\begin{itemize}
    \item store one graph file per source file in the project (requirements 1 and 2),
    \item adopt protobuf as the base format having a~small footprint (requirement 3), and being supported by many languages (requirement 5),
    \item schema definition contains a~flexible property map with any number of arguments (requirement 4).
\end{itemize}

Requirement 6 (conversion to other data formats) can easily be fulfilled by writing dedicated converters which is facilitated by utilizing Protocol Buffers, supported by all major languages\footnote{\url{https://protobuf.dev/reference/other/}}. 

The complete schema for the proposed format is presented in Listing \ref{lst:csgschema}. The format defines data structures for graph nodes, edges and location in the code, grouped together under one structure representing the entire source code file. Properties of the SCG, as defined in section \ref{semantic-code-graph-definition}, are mapped to individual fields of these structures. Additional dynamic node and edge properties can be placed in the dictionary structure \textit{properties} -- it is a single format extension point for any additional, often language-specific, properties.


The files are saved with the \texttt{.semanticgraph} extension.

\begin{lstlisting}[basicstyle=\footnotesize, caption=Protobuf Semantic Code Graph schema definition., label={lst:csgschema}]

syntax = "proto3";

message Location {
    string uri = 1;
    int32 startLine = 2;
    int32 startCharacter = 3;
    int32 endLine = 4;
    int32 endCharacter = 5;
}

message Edge {
    string to = 1;
    string type = 2;
    Location location = 3;
    map<string, string> properties = 4;
}

message GraphNode {
    string id = 1;
    string kind = 2;
    Location location = 3;
    map<string, string> properties = 4;
    string displayName = 5;
    repeated Edge edges = 6;
}

message SemanticGraphFile {
    string uri = 1;
    repeated GraphNode nodes = 2;
}
\end{lstlisting}

\section{Software Comprehension capabilities: comparison of SCG and other graph-based code representations}
\label{sec:scg-comparison}

In this section, we conduct a comparative analysis of various models as outlined in Section \ref{sec:related-work}, contrasting them with the SCG model within the context of software comprehension. We consider nine different code graph models in addition to the Semantic Code Graph (SCG): Control Flow Graph (CFG) \cite{allen1970control}, Program Dependence Graph (PDG) \cite{Ferrante1987}, Code Property Graph (CPG) \cite{Yamaguchi2014}, System Dependence Graph (SDG) \cite{horwitz1988interprocedural}, Java System Dependency Graph (JSysDG) \cite{Walkinshaw2003}, Call Graph (CG) \cite{Ryder1979}, Class Collaboration Network (CCN) \cite{Wheeldon2003}, Java Relationship Graphs (JRG) \cite{Arora2019}, and General Dependency Network (GDN) \cite{SAVIC2014}.

Our evaluation focuses on two key aspects: the models' suitability in representing crucial features from the perspective of code dependency software comprehension and their readiness for practical adoption in commercial settings. We break down the analysis into three categories. The first category addresses the general capabilities of each model, the second contains specific use cases vital for comprehending dependencies in modern programming languages, and the third concerns the readiness of these models for integration into tools and for conducting further research. While the evaluation criteria may have some subjectivity, we believe they effectively capture the essential functional and non-functional requirements necessary for a comprehensive and practical code structure representation model. A summarized comparison of the results is presented in Table \ref{tab:code-graph-comparisons}.

\begin{table*}[h!]
    \centering
    \fontsize{7pt}{10pt}\selectfont
    \begin{tabular}{p{7cm}|c|c|c|c|c|c|c|c|c|c|}
    & \multicolumn{5}{c|}{\textbf{Program execution focus}} & \multicolumn{5}{c|}{\textbf{Software dependency focus}} \\
    \cline{2-11}
                                                               & CFG & PDG & CPG & SDG & JSysDG & CG & CCN & JRG & GDN & SCG \\
    \cline{1-11}
    \multicolumn{11}{l}{\textbf{General model capabilities}} \\
    \cdashline{1-11}
    1. Language independent                                    &  +  &  +  &  +  &  +  &   -    & +  &  +  &  -  &   +  & + \\
    2. Captures interprocedural dependencies                   &  -  &  -  &  -  &  +  &   +    & +  &  +  &  +  &   +  & + \\
    3. Captures abstract code entities                         &  -  &  -  &  -  &  -  &   +    & -  &  +  &  +  &   +  & + \\
    4. Captures language specific dependencies                 &  -  &  -  &  -  &  -  &   +    & -  &  -  &  +  &   -  & + \\
    5. Preserves direct relation to the source code            &  -  &  -  &  -  &  -  &   -    & -  &  -  &  -  &   -  & + \\
    \cline{2-11}
                                                               &  1  &  1  &  1  &  2  &   3    & 2  &  3  &  3  &   3  & 5 \\
    \cline{1-11}
    \multicolumn{11}{l}{\textbf{Support for software comprehension use cases}} \\
    \cdashline{1-11}
    1. General focus on supporting software comprehension      &  -  &  -  &  -  &  -  &   -    & -  &  +  &  +  &   +  & + \\
    2. Can generate method Call Hierarchy                      &  -  &  -  &  -  &  +  &   +    & +  &  -  &  +  &   +  & + \\
    3. Can represent OOP entity relations                      &  -  &  -  &  -  &  -  &   +    & -  &  +  &  +  &   +  & + \\
    4. Can be used for 
       software structure analysis                             &  -  &  -  &  -  &  -  &   +    & -  &  +  &  +  &   +  & + \\
    5. Suitability for interactive source code visualizations  &  -  &  -  &  -  &  -  &   -    & -  &  -  &  -  &   -  & + \\
    \cline{2-11}
                                                               &  0  &  0  &  0  &  1  &   3    & 1  &  3  & 4  &   4  & 5 \\
    \cline{1-11}
    \multicolumn{11}{l}{\textbf{Tool adoption readiness}} \\
    \cdashline{1-11}
    1. Extraction program is 
       publicly available for Java language                    &  +  &  -  &  +  &  -  &   -    & +  &  +  &  -  &   -  & + \\
    2. The model defines data intermediate representation      &  -  &  -  &  +  &  -  &   -    & -  &  -  &  -  &   -  & + \\
    3. It is possible to persist extracted model               &  -  &  -  &  +  &  -  &   -    & -  &  -  &  -  &   -  & + \\
    4. It is possible to integrate 3rd party analysis tools    &  -  &  -  &  +  &  -  &   -    & -  &  -  &  -  &   -  & + \\
    \cline{2-11}
                                                               &  1  &  0  &  4  &  0  &   0    & 1  &  1  &  0  &   0  & 4 \\
    \cline{1-11}
    Readiness for practical software structure comprehension   & 14\% & 7\%&35\% & 21\%& 42\%   &28\%& 50\%& 50\%& 50\% &100\%\\
    \end{tabular}
    \caption{Different code graph representations and their capabilities. (CFG = Control Flow Graph; PDG = Program Dependence Graph; CPG = Code Property Graph; SDG = System Dependence Graph; JSysDG = Java System Dependency Graph; CG = Call Graph; CCN = Class Collaboration Network; GDN = General Dependency Network; SCG = Semantic Code Graph)}
    \label{tab:code-graph-comparisons}
\end{table*}

For the \textit{general model capabilities} category we specify the following traits:
\begin{enumerate}
    \item \textit{Language independent} -- capability of representing code written in different programming languages. This implies that the analysis of SCG data should function seamlessly irrespective of the programming language employed in a project (except in cases where the analysis relies on language-specific extended properties).
    \item \textit{Captures interprocedural dependencies} -- capability of representing dependencies between different procedures, so the model can be used to analyze the entire project.
    \item \textit{Captures abstract code entities} -- capability of representing entities which do not directly impact program execution (such as Java interfaces) yet are essential components of code structure.
    \item \textit{Captures language specific dependencies} -- ability to depict language-specific dependencies (such as inheritance for OOP languages, or \textit{extensions} from the Scala language).
    \item \textit{Preserves direct relation to the source code} -- ability to represent all entities found in the source code and point to concrete place in the source code where the entity or relation can be found (see Section \ref{sec:scg-requirements}).
\end{enumerate}

In this category, we evaluate the model's ability to comprehend the complete software landscape. Models solely focused on specific monolithic programs, such as CFG, PDG, or related CPG, do not support dependencies between different procedures and also fall short when it comes to handling higher-level dependencies among abstract units. Models that emphasize relationships between code entities prove to be a better fit for software comprehension. Most of these models are versatile and can be applied to analyze various programming languages, with the exception of JSysDG and JRG, which are specifically designed for the Java language. It is noteworthy that only SCG ensures a direct connection to the source code through the \textit{location} property. Besides SCG, JSysDG and JRG stand out as the most suitable representations of code, capable of illustrating even language-specific code dependencies within the entire program. Conversely, GDN is constructed on a generic extended concrete syntax tree (eCST), resulting in the loss of specific information, such as distinct node types. For instance, both \textit{enum} and \textit{class} are represented with the same node type referred to as \textit{CONCRETE\_UNIT\_DECL}.

To further assess the readiness of the model for software comprehension, we evaluate concrete use cases in the context of the given model:
\begin{enumerate}
    \item \textit{General focus on supporting software comprehension} -- is the model primarily focused on improving software comprehension rather than, for example, prioritizing program optimization.
    \item \textit{Can generate method Call Hierarchy} -- the ability to generate a call graph and its associated call hierarchy is a widely used representation when understanding software dependencies.
    \item \textit{Can represent OOP entities and relations} -- can the model effectively represent OOP elements such as objects and classes, as well as relationships such as polymorphism.
    \item \textit{Can be used for software structure analysis} -- is the model robust and detailed enough to support in-depth analysis for a better understanding of software structures.
    \item \textit{Can facilitate source code visualizations} -- when examining the graph, can users seamlessly navigate between the graph and the source code? The graph should closely align with the written source code and should be easy to comprehend without introducing unnecessary abstract graph nodes that may clutter the visualization.
\end{enumerate}

Supporting software comprehension is not the primary focus of statement-based program execution models. Consequently, Control Flow Graph (CFG), Program Dependence Graph (PDG), and Code Property Graph (CPG) models are limited to representing single procedures and are inadequate for various defined use cases. The System Dependency Graph (SDG) includes interprocedural call edges, which can assist in illustrating the Call Hierarchy. However, for understanding of software call chains, the Call Graph (CG) is the preferred choice, particularly popular among researchers. JSysDG builds upon the SDG and can accurately represent Object-Oriented Programming (OOP) features for the Java language. Class Collaboration Network (CCN) is also widely employed in software comprehension analysis. This model describes dependencies between classes, which are essential code entities in OOP. CCN is known for its simplicity, producing a manageable number of nodes what facilitates wide range of graph analysis. Models based on code entities offer more direct support for comprehending code dependencies and code structure, making them suitable representations for in-depth software structure analysis. However, interactive source code visualization requires an extremely accurate representation of the source code, complete with information on direct location pointers, a requirement uniquely fulfilled by the Semantic Code Graph (SCG) model.

Concerning the readiness for tool adoption, we assessed the following criteria:
\begin{enumerate}
    \item \textit{Extraction tool is publicly available for Java language} -- at least for the Java language (as an established and widely popular language) there is a publicly available tool to generate a representation of a source code. 
    \item \textit{The model defines data intermediate representation} -- a well-designed intermediate representation facilitates adding support for new languages; moreover, analysis tools relying on a~common intermediate representation are more versatile and language-agnostic. 
    \item \textit{It is possible to persist the extracted model} -- it is inefficient to extract the model from the source code each time an analysis is made. The ability to persist the model is therefore crucial.
    \item \textit{It is possible to integrate 3rd party analysis tools} -- there should be a means to export the extracted model for analysis with external tools commonly used by researchers, such as Jupyter Notebooks or the Gephi visualization software.
\end{enumerate}

In Section \ref{sec:related-work}, we discussed various graph source code models, whose utility on both theoretical and practical levels has been demonstrated by developing experimental extraction tools. Unfortunately, few of these tools are publicly shared, making it challenging for other researchers to reproduce results, or enhance existing implementations, and conduct further analyses. For Java, the Control Flow Graph (CFG) and Call Graph (CG) can be extracted using the \textit{ScootUp}\footnote{\url{https://github.com/soot-oss/SootUp/}} tool \cite{lam2011soot}. The Code Property Graph (CPG) can be generated using the \textit{joern}\footnote{\url{https://github.com/joernio/joern}} platform, which extracts CPGs for various languages (with stable implementation for C/C++ and experimental for Java/Kotlin/Javascript/Python) into an embedded database and provides a custom Scala DSL for database queries. CPG provides a specification of an intermediate representation\footnote{\url{https://cpg.joern.io/}}, and it supports the storage of extracted CPG models for future use. Furthermore, it offers seamless integration with third-party software through direct Scala API or export to formats such as GraphML, DOT, or other graph formats. The \textit{joern} platform generates  Program Dependency Graphs (PDG) and Control Flow Graphs (CFG) for C/C++ languages but it has a potential to support other languages in the future. Shu et al. \cite{Shu_2013} describe the \textit{JavaPDG} tool capable of extracting Java PDG graphs. However, the link to the tool is outdated, and the actual implementation is currently unavailable. In terms of entity relations, basic networks, such as Class Collaboration Networks, can be extracted using \textit{DependencyFinder}. Extraction programs for PDG, JSysDG, JRG, and GDN for Java language do not currently exist. For the Semantic Code Graph (SCG), we created a command line tool called \textit{scg-cli}\footnote{\url{https://github.com/VirtusLab/scg-cli}}. This tool is described in more detail in section \ref{subsec:scg-cli-intro}. 

In summary, the SCG model fulfills all proposed criteria, while GDN, JRG, and CCN achieved the readiness for software comprehension score of $50\%$. The main shortcomings of these models were related to the lack of direct code relationships, readiness for code visualization, and tool adoption capabilities. Despite its primary focus on comprehensive program execution analysis, JSysDG achieved a  score of 42\%. Its limitations were mainly attributed to practical constraints within the model and the absence of a comprehensive extraction program. CPG exhibited robust tooling support, primarily due to its extensive utilization in identifying vulnerabilities within C/C++ programs. Although CPG also offers support for the Java language, it remains in an experimental phase, still lacking in terms of feature completeness and model implementation precision. On the other hand, our proposed SCG model maintains excellent software comprehension capabilities with strong practical support for additional tool integration.

\section{Empirical Software Comprehension with Semantic Code Graph}
\label{sec:scg-software-comprehension}



In this section, we conduct several experimental software comprehension activities on selected projects, in order to answer research questions stated earlier:
\begin{itemize}
    \item \textbf{RQ1}: Does the SCG model enhance software comprehension capabilities in comparison with the Class Collaboration Network and the Call Graph models?
    \item \textbf{RQ2}: Does SCG-based data analysis enable actionable software comprehension insights?
\end{itemize}


Consequently, we extract three graph models -- the SCG, the CCN and the CG -- from eleven well-known Java and Scala open source projects and analyze the collected data using various techniques, such as complex network analysis \cite{Newman2018networks} and the split-apply-combine data analysis methods \cite{Wickham2011split}.

The eleven open source projects used in this study are as follows:
\begin{itemize}
    \item retrofit\footnote{\url{https://github.com/square/retrofit}} -- a type-safe HTTP client for Android and Java,
    \item commons-io\footnote{\url{https://github.com/apache/commons-io}} -- the Apache Commons IO library containing utility classes, stream implementations, file filters, file comparators, endian transformation classes,
    \item playframework\footnote{\url{https://github.com/playframework/playframework}} -- high velocity web framework for building web applications with Java \& Scala,
    \item metals\footnote{\url{https://github.com/scalameta/metals}} -- Scala language server with rich IDE features,
    \item glide\footnote{\url{https://github.com/bumptech/glide}} -- media management and image loading framework for Android that wraps media decoding, memory and disk caching, and resource pooling into a simple and easy to use interface,
    \item vert.x\footnote{\url{https://github.com/eclipse-vertx/vert.x}} -- toolkit to build reactive applications on JVM; Vert.x Core library containing low-level functionalities including support for HTTP, TCP and file system access,
    \item RxJava\footnote{\url{https://github.com/ReactiveX/RxJava}} -- Reactive Extensions for the JVM; a library for composing asynchronous and event-based programs by using observable sequences,
    \item dubbo\footnote{\url{https://github.com/apache/dubbo}} -- Apache Dubbo is a high-performance, Java-based open-source RPC framework,
    \item spring-boot\footnote{\url{https://github.com/spring-projects/spring-boot}} -- popular library to help with creating Spring applications,
    \item Akka\footnote{\url{https://github.com/akka/akka}} -- a toolkit for building highly concurrent, distributed, and resilient message-driven applications for Java and Scala, 
    \item Spark\footnote{\url{https://github.com/apache/spark}} -- a well-known engine for large-scale data analytics.
\end{itemize}

The selected projects are very popular among programmers -- nine out of eleven have more than 10k stars on GitHub repository. 

We have conducted the experiments with the help of the \textit{scg-cli} tool. From a practitioner's standpoint, our objective is to provide developers with an introduction to the SCG software. To facilitate reproducibility of our results, we have published all the extracted SCG data used in this study in the form of protobuf files, available in the \textit{data} folder of the \textit{scgi-cli} github repository\footnote{https://github.com/VirtusLab/scg-cli/tree/main/data}.

For each project, we build and analyze the full SCG graph, and then extract the Class Collaboration Network and the Call Graph from the SCG data. We have chosen these two models in addition to SCG, as both the Call Graph \cite{Ryder1979, LaToza2011, Guo2013, Shah2016, Falci2017, Alanzi2021} and the Class Collaboration Network \cite{savic2012community, Meyer2015, Pan2018, Li2021, Du2021} are models that are popular among researchers and have been used for software comprehension. 

\begin{figure}[!h]
\centering
\begin{subfigure}[b]{0.4\textwidth}
  \centering
  \begin{lstlisting}[basicstyle=\footnotesize]
class D
class C  
trait A 
class B extends A {
  var d: D
  def bar(c: C): D = d
}
  \end{lstlisting}
  \caption{Scala inheritance and aggregation example.}
\end{subfigure}%

\begin{subfigure}[b]{0.4\textwidth}
  \centering
  \includegraphics[width=150px]{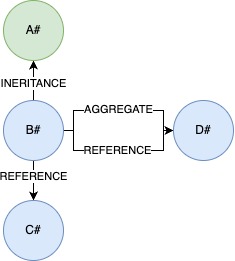}
  \caption{Class Collaboration Network example visualization.}
\end{subfigure}
\caption{Class Collaboration Network example with \textit{INHERITANCE} dependencies between \textit{class B} and \textit{class A}, \textit{AGGREGATION} representing variable \textit{D} and \textit{REFERENCE} representing method param and return type coupling.}
\label{fig:ccn-example}
\end{figure}

Our definition of the Class Collaboration Network is akin to the one presented by Wheeldon and Counsell \cite{Wheeldon2003}. Figure \ref{fig:ccn-example} illustrates the concept of CCN with three distinct class coupling types: inheritance, aggregation, and reference through parameter and return types. A CCN model is extracted from the SCG data by iteratively traversing relevant nodes, and creating a new graph comprising nodes of types \textit{OBJECT}, \textit{CLASS}, \textit{TRAIT}, \textit{INTERFACE}, and their associated edges: \textit{INHERITANCE}, \textit{AGGREGATION}, and \textit{REFERENCE}.

The Call Grah is obtained as a~SCG subgraph with edges limited to type \textit{CALL} and vertices of types \textit{METHOD}, \textit{CONSTRUCTOR}, \textit{VALUE} and \textit{VARIABLE}. 

The remainder of this section is organized as follows. Section \ref{subsec:scg-cli-intro} briefly describes the \textit{scg-cli} tool. Section \ref{sec:project-overview} presents the first case of software comprehension, where we show how simple analysis of project metrics, followed by more complex software mining, can lead to interesting insights. Section \ref{sec:project-structure-comprehension} provides an examination of the projects' structural characteristics. Section \ref{sec:project-critical-entities} continues with identification of critical code entities which typically serve as excellent starting points for comprehending the source code. Exploration of project structure can be further enhanced using interactive visualization with the Graph Buddy tool, briefly described in Section \ref{sec:reachability}. Section \ref{sec:finding-similarities} demonstrates the capabilities of custom software mining analysis, specifically the identification of code similarities, made feasible by applying data analysis methods to the the rich SCG model. Finally, we conclude the empirical evaluation by addressing our research questions in Section \ref{sec:answering-research-questions}.

\subsection{The scg-cli tool}
\label{subsec:scg-cli-intro}
To facilitate software comprehension using the SCG model, we have created and open-sourced the \textit{scg-cli}\footnote{\url{https://github.com/VirtusLab/scg-cli}} tool. Ready-to-use binaries can be downloaded from the github release page\footnote{\url{https://github.com/VirtusLab/scg-cli/releases}}.

The \textit{scg-cli} tool offers several functionalities, including extraction of the SCG data from Java projects, storage of the SCG data in a compressed protobuf format with a publicly known schema, and exporting to popular graph formats. It also provides a~Jupyter environment wherein the SCG data can be analyzed using Python Pandas dataframes. Moreover, \textit{scg-cli} can generate CCN and CG graphs from the collected SCG data using command arguments \textit{--graph CCN,CG}. 

Additionally, the \textit{scg-cli} command-line tool directly empowers software comprehension through following capabilities:
\begin{itemize}
    \item Generating project summary to give the programmer a~high level overview of the project.
    \item Finding critical entities in the source code that are usually good starting points for reading the project.
    \item Discovering and suggesting software partitioning to expose module boundaries or support refactoring efforts.
\end{itemize}

Software comprehension activities, such as identifying critical entities, software partitioning, or exporting results to \textit{.GDF} or \textit{graphml} output formats, can operate on all three supported software networks: SCG, CCN, and CG. All aforementioned capabilities were extensively utilized in our empirical research, which demonstrates the usefulness of this tool, and facilitates reproducibility of the presented results. 

\subsection{Project overview}
\label{sec:project-overview}

\begin{figure*}[!htb]
    \centering
    \includegraphics[width=0.9\textwidth]{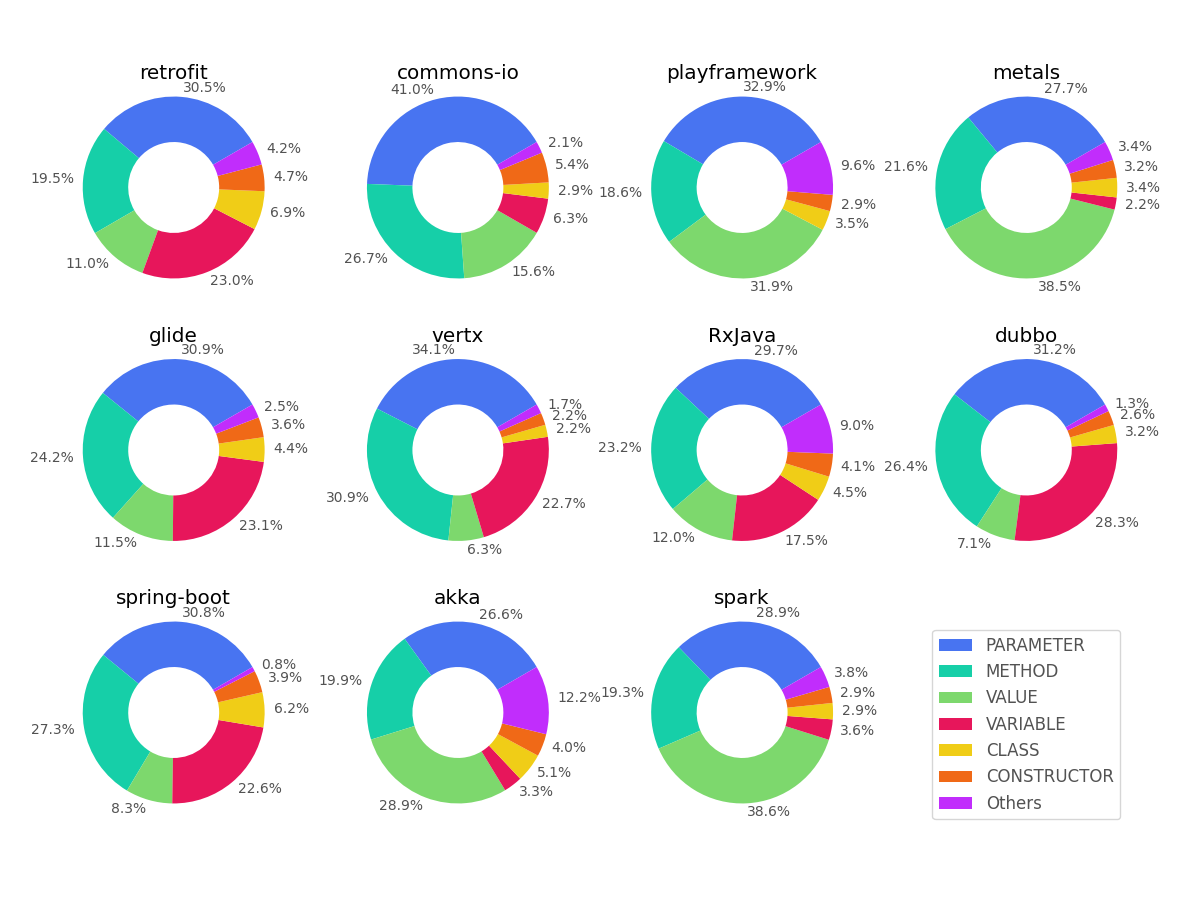}
    \caption{Top 6 SCG nodes distribution in analyzed projects.}
    \label{fig:nodes-distribution}
\end{figure*}

\begin{figure*}[!htb]
    \centering
    \includegraphics[width=0.9\textwidth]{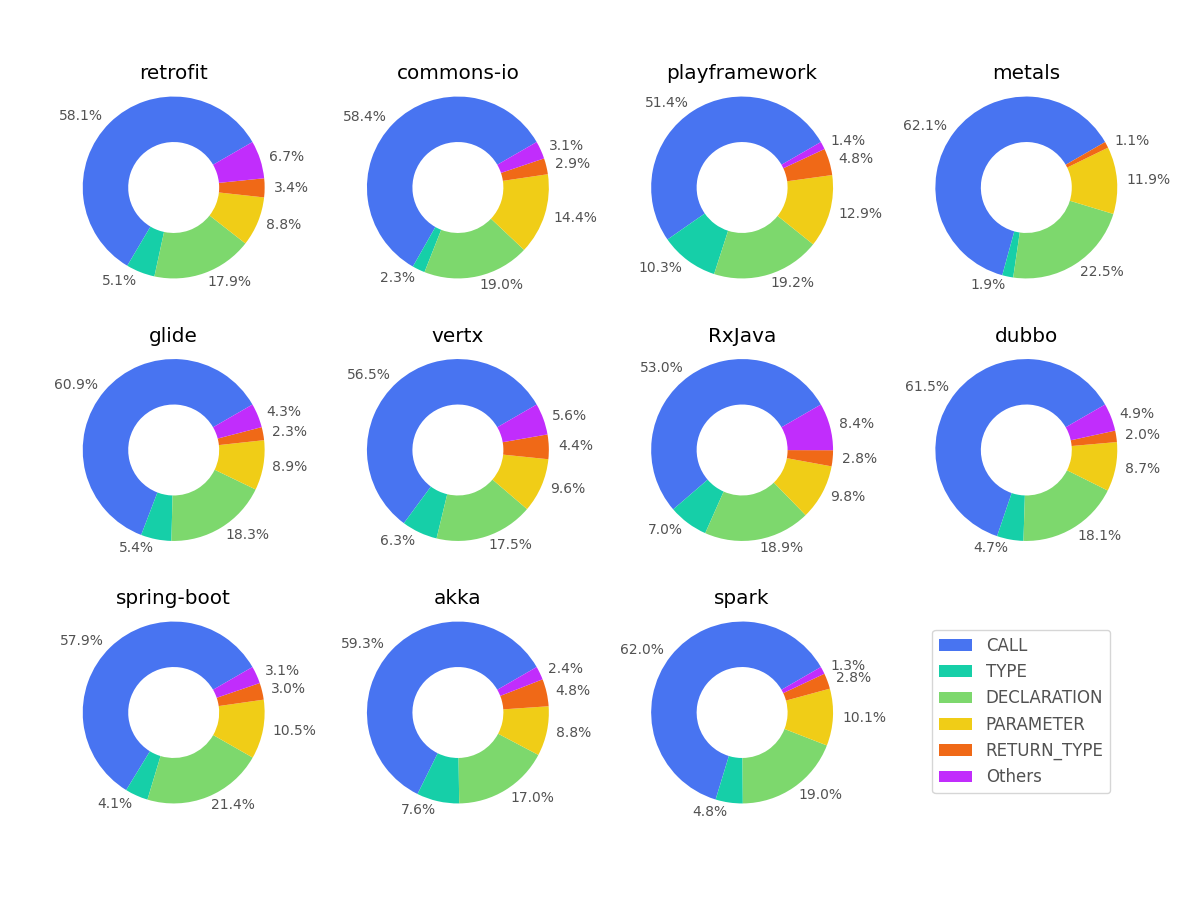}
   \caption{Top 5 SCG edges distribution in analyzed projects.}
    \label{fig:edges-distribution}
\end{figure*}

We start the software comprehension experiment with analysis of the simplest properties: the distribution of nodes (Fig. \ref{fig:nodes-distribution}) and edges (Fig. \ref{fig:edges-distribution}) in the SCG graphs of the investigated projects. Looking at the node distribution, we can immediately observe that all Java projects, except for \textit{commons-io}, have a relatively high fraction of variables, reaching 22.5\% for \textit{spring-boot}, compared to only 6.3\% in the case of \textit{common-io}, and less than 4\% for Scala projects. Since excessive use of variables can be harmful, we decide to use \textit{software mining} techniques to investigate this further. First, for the \textit{spring-boot} project, we find top files in terms of the number of variables in the project. With the help of the \textit{scg-cli} tool, we export the project SCG metadata into Jupyter notebook for closer inspection. The notebook provides a python environment with a~helper API to read the SCG protobuf data into Pandas\footnote{\url{https://pandas.pydata.org}} data frames. Next, we use a~standard split-apply-combine data analysis strategy, as shown in Listing \ref{lst:spring-boot-pandas}. With this short program we can find 11807 local variables and discover the files with the highest number of them, e.g., \textit{ConfigurationPropertyName.java}, \textit{PaketoBuilderTests.java}, \textit{JarIntegrationTests.java}, \textit{OnBeanCondition.java} and \textit{TomcatWebServerFactoryCustomizer.java}. 

\begin{lstlisting}[basicstyle=\footnotesize, language=Python, caption=Dataframe-based analysis of the spring-boot SCG data, label={lst:spring-boot-pandas}]
import scg

scg_files = scg.read_scg("data/spring-boot")
df = scg.create_nodes_df(scg_files)
variables_df = df[df['kind']=="VARIABLE"]

variables_df.filter(items=['id', 'isLocal']) \
    .groupby('isLocal').count()
#             id
# isLocal       
# false     2639
# true     11807

variables_df[variables_df['isLocal']==True] \
  .filter(items=['id', 'file']) \
  .groupby('file') \
  .agg(count=('id', 'count')) \
  .sort_values(by='count', ascending=False)
#                                          count
# file                                              
# ConfigurationPropertyName.java             102
# PaketoBuilderTests.java                    101
# JarIntegrationTests.java                    86
# OnBeanCondition.java                        80
# TomcatWebServerFactoryCustomizer.java       72

variables_df[variables_df['isLocal']==False] \
  .filter(items=['id', 'file']) \
  .groupby('file') \
  .agg(count=('id', 'count')) \
  .sort_values(by='count', ascending=False)
#                                        count
# file                                                
# ServerProperties.java                    104
# KafkaProperties.java                      76
# RabbitProperties.java                     68
# FlywayProperties.java                     60
# WebProperties.java                        31  
\end{lstlisting}

Looking at the source code of \textit{ConfigurationPropertyName.java}, we can find many examples of mutable variables as presented in Listing \ref{lst:spring-boot-local-variables} (e.g., \texttt{int start}) or immutable variables missing the \texttt{final} modifier (e.g., \texttt{int length}). It can be observed that many cases of the latter particularly contribute to the large number of variables. In comparison, the \textit{common-io} project correctly uses the \texttt{final} keyword for each immutable local variable (see Listing \ref{lst:commons-io-final-keyword-application}), a~good practice in the Java code, effectively transforming such variables into immutable \textit{VALUEs}. Consequently, we can conclude that the quality of the source code of the \textit{spring-boot} project could be improved by eliminating the excessive number of local mutable variables. 
\begin{lstlisting}[basicstyle=\footnotesize, language=Java, caption=Example of local variables' definitions in spring-boot \textit{ConfigurationPropertyName.java} file, label={lst:spring-boot-local-variables}]
Elements parse(Function<CharSequence, 
               CharSequence> valueProcessor) {
    int length = this.source.length();
    int openBracketCount = 0;
    int start = 0;
    ElementType type = ElementType.EMPTY;
    ...
}
\end{lstlisting}

\begin{lstlisting}[basicstyle=\footnotesize, language=Java, caption=Example of final local variables definition in \textit{commons-io} project, label={lst:commons-io-final-keyword-application}]
public static void copy(
        final String input,
        final OutputStream output,
        final String encoding)
            throws IOException {
    final StringReader in = 
        new StringReader(input);
    final OutputStreamWriter out = 
        new OutputStreamWriter(output, encoding);
    copy(in, out);
    out.flush();
}
\end{lstlisting}

A different case of using variables that we have found are non-local variables present in \textit{spring-boot} property files, such as \textit{ServerProperties.java}, \textit{KafkaProperties.java}, \textit{RabbitProperties.java}, \textit{FlywayProperties.java}. These files contain different configuration settings, such as the defined \texttt{port} field presented in Listing \ref{lst:spring-boot-global-variables}, which are accessed through global accessors. While this has been a standard Java approach for classes representing some internal state, currently the recommended practice is to introduce specialized constructors and add \texttt{final} modifiers for global private fields to eliminate accidental mutability. Performing such refactoring is another example of recommendation that we can formulate in order to increase the quality of the project source code. 

\begin{lstlisting}[basicstyle=\footnotesize, language=Java, caption=Example of global variable definition in spring-boot \textit{ServerProperties.java} file with accessors methods, label={lst:spring-boot-global-variables}]
/**
* Server HTTP port.
*/
private Integer port;

public Integer getPort() {
    return this.port;
}

public void setPort(Integer port) {
    this.port = port;
}
\end{lstlisting}


\textit{\textbf{Conclusions.}} The SCG represents a variety of code entities and their relations, offering a much more detailed project overview compared to CCN or CG models. Visualization of the percentage distribution of main code entities and relations (Fig. \ref{fig:nodes-distribution} and Fig. \ref{fig:edges-distribution}) has led us not only to identify an excessive number of variables in some Java projects, but also to distinguish between local and global variables. This enabled us to easily identify the most problematic files, draw conclusions, and suggest actionable insights. Thanks to the level of detail provided by the SCG model, where nodes can be augmented with additional contextual information (such as entity LOC size and entity code complexity), comprehensive overviews can be presented. 

\subsection{Project structure comprehension}
\label{sec:project-structure-comprehension}
In this example, we demonstrate how project structure discovery can be performed by looking at various graph metrics, described in Table \ref{tab:scg-measured-properties}. For each project, we computed all metrics for the full SCG, the Call Graph and the Class Collaboration Network. The results are presented in Table \ref{tab:scg-project-summary}.

\begin{table*}[h!]
    \centering
    \fontsize{9pt}{8pt}\selectfont
    \begin{tabular}{p{3cm}|p{13cm}}
        Property & Explanation \\
        

         \cline{1-2}
         \multirow{2}{3cm}{Density ($D$)} & $$|D|=\frac{|E|}{|V|(|V|-1)}$$ \\
         & Density of the graph -- measure of how closely connected the graph is. High SCG density can indicate higher code dependency complexity.  $D \in [0,1]$\\
         
         \cline{1-2}
         \multirow{2}{3cm}{Average Node Degree ($A_{D}$)} & $$A_{D} = \frac{|E|}{|V|}$$ \\
         & High average degree of nodes, especially combined with high density, suggests high project coupling; on the contrary, high node average degree with low density might imply a significant number of hubs. \\

         \cline{1-2}
         \multirow{2}{3cm}{Standard Deviation of incoming/outgoing node degree distribution ($\sigma_{ID}$/$\sigma_{OD}$)} & $$\sigma = \sqrt{\frac{1}{|V|} \sum_{i=1}^{|V|} (d_i - A_{D})^2}$$ 
         Where:
            \begin{itemize}
              \item $|V|$ is number of nodes
              \item $d_i$ is an incoming or outgoing degree of a node $i$
            \end{itemize}
         \\
         & A high standard deviation in the node degree distribution indicates significant variability in node degrees. This may suggest that certain entities exhibit considerably higher incoming or outgoing degrees than the average. Consequently, a high standard deviation might indicate greater code complexity and coupling. \\

         \cline{1-2}
         \multirow{2}{3cm}{Index of Dispersion of incoming/outgoing node degree distribution ($IoD_{ID}$/$IoD_{OD}$)} & $$\text{IoD} = \frac{\sigma^2}{A_{D}}$$ 
         Where:
            \begin{itemize}
              \item $\sigma^2$ is the variance of the node degrees (incoming or outgoing)
            \end{itemize}
         \\
         &  Related to Standard Deviation. If the $IoD$ is greater than 1, it indicates that the node degrees are more variable (dispersed) than what one would expect from a Poisson distribution. A high $IoD$ suggests that the network may possess a scale-free structure, characterized by a few highly connected nodes (hubs) and numerous nodes with lower degrees. \\

         \cline{1-2}
         \multirow{2}{3cm}{Average Clustering Coefficient (ACC)} & $$ACC = \frac{1}{|V|} \sum_{i=1}^{|V|} \frac{2T_i}{d_i(d_i - 1)}$$ 
         Where:
         \begin{itemize}
             \item \(T_i\) represents the number of triangles (closed triplets) that node \(i\) is part of.
             \item \(d_i\) represents the degree of node \(i\)
         \end{itemize}
         \\
         & The average clustering coefficient provides an average measure of how tightly interconnected nodes are in local neighborhoods across the entire network. $ACC \in [0, 1]$\\
         
         \cline{1-2}
         \multirow{2}{3cm}{Global Clustering Coefficient (GCC)} & $$GCC = 3 \times \frac{number\_of\_triangles}{number\_of\_triplets}$$ \\
         & High value indicates that graph tends to create interconnected clusters, low value indicates the network being sparse or fragmented. For SCG high GCC value can indicate well-modularized software with high module cohesion and low coupling. $GCC \in [0, 1]$\\

         \cline{1-2}
         \multirow{2}{3cm}{Degree Assortativity Coefficient (DAC)} & 
         $$DAC=\frac{\sum_{i,j}(A_{ij} - d_ik_j/2m)d_id_j}{\sum_{i,j}(A_{i,j}d_i^2 - \frac{d_id_j}{2m}d_id_j)} $$
         where \cite{Newman2018networks}:
         \begin{itemize}
            \item $A$ is the adjacency matrix
            \item $d_i$ is the degree of the node $i$
            \item $m$ is the total number of edges
         \end{itemize} \\
         & Positive value indicates that nodes with the same degree tends to be connected, negative value suggest mostly connections between nodes of different degree, value close to zero suggest random network. For SCG low DAC can suggest additional tendency of the graph towards clustering and establishing hubs (connection between nodes with different degree). $DAC \in [-1, 1]$\\
        
         \cline{1-2}
    \end{tabular}
    \caption{Metrics calculated for the SCG, CCN and CG software networks.}
    \label{tab:scg-measured-properties}
\end{table*}

\begin{table*}[h!]
\centering
\begin{footnotesize}
\setlength{\tabcolsep}{0.4em}
\begin{tabular}{r|l|r|r|r|r|r|r|r|r|r|r|r|r|r|} 

Name & Version & \#LOC & $|V|$ & $\frac{\#LOC}{|V|}$ & $|E|$ & $D$ & $A_{D}$ & $\sigma_{ID}$ & $IoD_{ID} $ & $\sigma_{OD}$ & $IoD_{OD}$ & ACC & GCC & DAC \\
\cline{1-15}
\multicolumn{15}{l|}{Semantic Code Graph} \\
\cdashline{1-15}
retrofit & 2.9.0 & 13676 & 3035 & 4.51  & 10137 & 0.00110 & 3.3 & 6.2 & 11.7 & 13.8 & 57.4 & 0.16 & 0.03 & -0.05 \\
commons-io & 2.12.0 & 46526 & 7418 & 6.27  & 21257 & 0.00039 & 2.9 & 3.3 & 3.7 & 7.2 & 17.8 & 0.13 & 0.07 & -0.23 \\
playframework & 2.8.19 & 57592 & 23058 & 2.50  & 67080 & 0.00013 & 2.9 & 9.1 & 28.6 & 10.5 & 38.0 & 0.10 & 0.03 & -0.08 \\
metals & 0.10.3 & 58172 & 19511 & 2.98  & 53347 & 0.00014 & 2.7 & 3.2 & 3.8 & 10.5 & 40.7 & 0.08 & 0.03 & -0.14 \\
glide & 4.5.11 & 65397 & 15459 & 4.23  & 53211 & 0.00022 & 3.4 & 5.3 & 8.3 & 8.3 & 20.1 & 0.15 & 0.08 & -0.27 \\
vert.x & 4.4.4 & 102189 & 24498 & 4.17  & 86664 & 0.00014 & 3.5 & 12.5 & 44.1 & 10.8 & 33.2 & 0.13 & 0.03 & 0.17 \\
RxJava & 3.1.6 & 188189 & 37729 & 4.99  & 118788 & 0.00008 & 3.1 & 11.9 & 45.2 & 7.7 & 19.0 & 0.12 & 0.02 & -0.06 \\
dubbo & 3.2.4 & 247943 & 57601 & 4.30  & 203693 & 0.00006 & 3.5 & 11.2 & 35.5 & 9.3 & 24.5 & 0.14 & 0.04 & -0.04 \\
spring-boot & 2.7.5 & 330424 & 64190 & 5.15  & 186816 & 0.00005 & 2.9 & 4.0 & 5.4 & 6.0 & 12.3 & 0.16 & 0.09 & -0.12 \\
akka & 2.7.0 & 337559 & 122355 & 2.76  & 436691 & 0.00003 & 3.6 & 17.8 & 88.9 & 20.3 & 115.9 & 0.12 & 0.01 & -0.05 \\
akka (without tests) & 2.7.0 & 234445 & 90195 & 2.60  & 295957 & 0.00004 & 3.3 & 13.4 & 54.5 & 11.3 & 38.9 & 0.12 & 0.02 & -0.08 \\
spark & 3.3.0 & 527149 & 164772 & 3.20  & 561314 & 0.00002 & 3.4 & 16.9 & 84.0 & 16.7 & 81.9 & 0.09 & 0.01 & -0.03 \\

\cline{1-15}
\multicolumn{15}{l|}{Class Collaboration Network} \\
\cdashline{1-15}
retrofit & 2.9.0 & 13676 & 183 & 74.73  & 473 & 0.01420 & 2.6 & 12.2 & 57.5 & 2.0 & 1.5 & 0.23 & 0.08 & -0.61 \\
commons-io & 2.12.0 & 46526 & 174 & 267.39  & 438 & 0.01455 & 2.5 & 8.8 & 30.9 & 4.9 & 9.7 & 0.09 & 0.09 & -0.22 \\
playframework & 2.8.19 & 57592 & 1208 & 47.68  & 4935 & 0.00338 & 4.1 & 29.3 & 209.9 & 19.2 & 89.9 & 0.09 & 0.02 & 0.10 \\
metals & 0.10.3 & 58172 & 680 & 85.55  & 1427 & 0.00309 & 2.1 & 4.6 & 10.1 & 3.2 & 4.8 & 0.05 & 0.10 & -0.61 \\
glide & 4.5.11 & 65397 & 707 & 92.50  & 2597 & 0.00520 & 3.7 & 11.6 & 36.7 & 5.5 & 8.4 & 0.16 & 0.15 & -0.42 \\
vert.x & 4.4.4 & 102189 & 600 & 170.32  & 4960 & 0.01380 & 8.3 & 39.3 & 186.6 & 13.4 & 21.8 & 0.18 & 0.46 & -0.30 \\
RxJava & 3.1.6 & 188189 & 1564 & 120.33  & 6787 & 0.00278 & 4.3 & 36.1 & 299.7 & 11.0 & 28.1 & 0.16 & 0.84 & -0.40 \\
dubbo & 3.2.4 & 247943 & 2030 & 122.14  & 8641 & 0.00210 & 4.3 & 33.1 & 256.6 & 5.8 & 7.8 & 0.15 & 0.03 & -0.17 \\
spring-boot & 2.7.5 & 330424 & 3199 & 103.29  & 7096 & 0.00069 & 2.2 & 6.1 & 16.8 & 3.5 & 5.5 & 0.10 & 0.09 & -0.14 \\
akka & 2.7.0 & 337559 & 8107 & 41.64  & 27250 & 0.00041 & 3.4 & 30.4 & 274.4 & 12.9 & 49.9 & 0.06 & 0.19 & -0.09 \\
akka (without tests) & 2.7.0 & 234445 & 5477 & 42.81  & 21254 & 0.00071 & 3.9 & 31.4 & 253.7 & 15.5 & 62.2 & 0.07 & 0.25 & -0.05 \\
spark & 3.3.0 & 527149 & 6461 & 81.59  & 31742 & 0.00076 & 4.9 & 54.1 & 594.8 & 13.6 & 37.6 & 0.12 & 0.06 & -0.19 \\

\cline{1-15}
\multicolumn{15}{l|}{Call Graph} \\
\cdashline{1-15}
retrofit & 2.9.0 & 13676 & 1677 & 8.16  & 4151 & 0.00148 & 2.5 & 4.1 & 6.9 & 11.4 & 52.1 & 0.13 & 0.03 & -0.06 \\
commons-io & 2.12.0 & 46526 & 3604 & 12.91  & 7232 & 0.00056 & 2.0 & 3.2 & 5.0 & 4.9 & 12.0 & 0.08 & 0.06 & -0.22 \\
playframework & 2.8.19 & 57592 & 9469 & 6.08  & 16750 & 0.00019 & 1.8 & 5.9 & 19.9 & 6.5 & 24.0 & 0.01 & 0.01 & -0.05 \\
metals & 0.10.3 & 58172 & 10900 & 5.34  & 20638 & 0.00017 & 1.9 & 2.9 & 4.4 & 7.9 & 33.4 & 0.02 & 0.01 & -0.11 \\
glide & 4.5.11 & 65397 & 8863 & 7.38  & 24807 & 0.00032 & 2.8 & 5.0 & 8.9 & 7.2 & 18.4 & 0.12 & 0.06 & -0.22 \\
vert.x & 4.4.4 & 102189 & 13431 & 7.61  & 33828 & 0.00019 & 2.5 & 5.4 & 11.6 & 6.4 & 16.5 & 0.10 & 0.05 & -0.22 \\
RxJava & 3.1.6 & 188189 & 19348 & 9.73  & 43800 & 0.00012 & 2.3 & 5.3 & 12.5 & 5.9 & 15.2 & 0.10 & 0.04 & -0.14 \\
dubbo & 3.2.4 & 247943 & 33038 & 7.50  & 91100 & 0.00008 & 2.8 & 7.5 & 20.2 & 7.4 & 19.7 & 0.12 & 0.04 & -0.14 \\
spring-boot & 2.7.5 & 330424 & 36803 & 8.98  & 73727 & 0.00005 & 2.0 & 3.8 & 7.3 & 4.4 & 9.8 & 0.10 & 0.07 & 0.02 \\
akka & 2.7.0 & 337559 & 47621 & 7.09  & 107184 & 0.00005 & 2.3 & 12.5 & 69.8 & 7.0 & 21.8 & 0.01 & 0.01 & -0.05 \\
akka (without tests) & 2.7.0 & 234445 & 37916 & 6.18  & 87809 & 0.00006 & 2.3 & 11.9 & 61.5 & 7.3 & 23.3 & 0.01 & 0.01 & -0.05 \\
spark & 3.3.0 & 527149 & 90285 & 5.84  & 208191 & 0.00003 & 2.3 & 10.1 & 44.1 & 9.7 & 40.9 & 0.01 & 0.01 & -0.08 \\

\end{tabular}
\end{footnotesize}
\caption{Project characteristics extracted from the Semantic Code Graph, Class Collaboration Network and the Call Graph ($D$ = Density, $A_{OD}$ = Average out-degree, $A_{ID}$ = Average in-degree, $\sigma$ = Standard Degree Distribution Deviation, IoD = Degree Distribution Index of Dispersion, ACC = Average Clustering Coefficient, GCC = Global Clustering Coefficient, DAC = Degree Assortativity Coefficient)}
\label{tab:scg-project-summary}
\end{table*}

The metrics show that for Java projects, there are on average more lines of code per graph node (code entity) than in Scala projects. This is expected as the Scala language is much more concise than Java. Nevertheless, the difference is quite significant with around 1.7 times as many code lines per node for Java projects. These metrics suggest that the Scala language can greatly contribute to lower code verbosity.

All projects have low \textit{density}, although the SCG graph for the \textit{retrofit} project is around 55 times denser than for the \textit{spark} project. This metric is much higher also for CCN and CG models, possibly indicating a~higher overall coupling level.

\begin{figure*}[h!]
    \centering
    \includegraphics[width=0.9\textwidth]{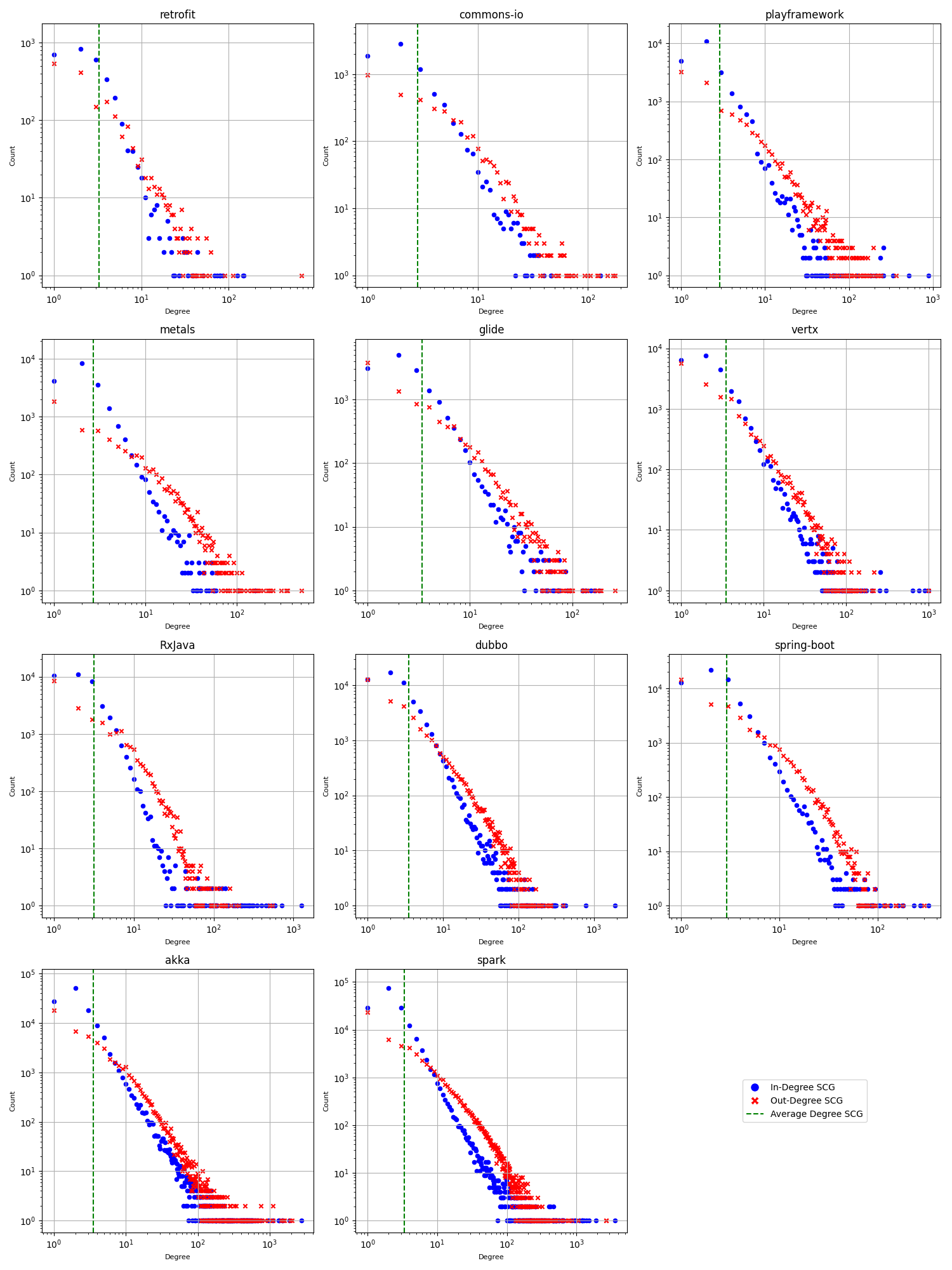}
    \caption{Nodes' incoming and outgoing degree distribution for SCG model extracted for each analyzed project.}
    \label{fig:distribution-scg}
\end{figure*}

The incoming and outgoing degree distribution of SCG nodes, as presented in Figure \ref{fig:distribution-scg}, follows a power-law distribution for each project, indicating that the network exhibits characteristics of a scale-free network \cite{myers2003software, hyland2006scale}. This is further supported by the low average node degree for each network, combined with a high standard deviation ($\sigma$) and a high index of dispersion ($IoD$) for both incoming and outgoing degree distributions. A scale-free network indicates the likely existence of highly connected code entities (hubs), which may represent core functionalities. Changes, removals, or vulnerabilities in relation to random low-level nodes are likely to have minimal effects on the software. Conversely, altering hubs can significantly impact the majority of the software in terms of structure, correctness and program execution performance. Upon examining all eleven analyzed projects, it becomes evident that the characteristics of node distribution are not correlated with either the project size or the project's programming language. Additionally, although the average degree for each project is similar (approximately 3), the standard deviation and index of dispersion can vary significantly. A particularly high index of dispersion may signal the presence of outlier nodes with exceptionally high degree levels.

\begin{figure*}[h!]
    \centering
    \includegraphics[width=0.9\textwidth]{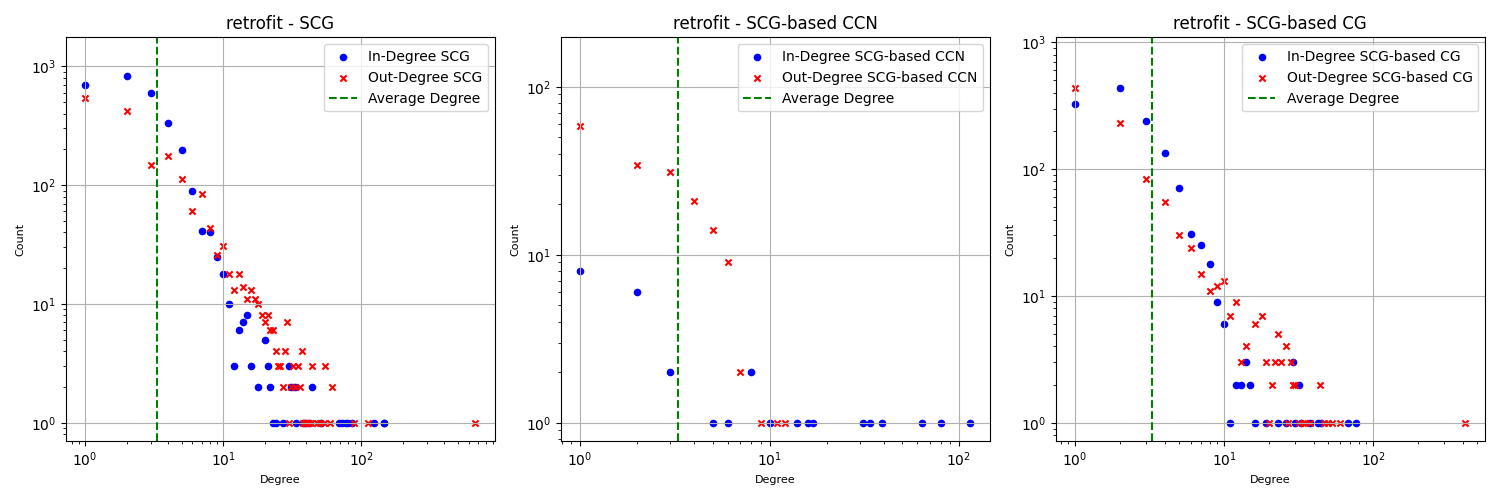}
    \caption{The incoming and outgoing degree distribution of nodes in the SCG, CCN, and CG models was extracted for the \textit{retrofit} project, with a focus on identifying outlier nodes with high outgoing degrees in the SCG and CG networks.}
    \label{fig:distribution-retrofit}
\end{figure*}

For example, the \textit{Retrofit} project has an exceptionally high index of dispersion (third highest) for the outgoing node degree, observed both in SCG and CG networks, with values of 57.4 and 52.1, respectively. At the same time, this parameter is very small for the CCN network (the lowest), which is different to what we can observe for other projects. When examining the node degree distribution of the \textit{Retrofit} project, as presented in Figure \ref{fig:distribution-retrofit}, we observe an outlier node which corresponds to \textit{Builder\#parseParameterAnnotation}, a~455 lines-of-code method that has an extremely high outgoing node degree of 674. Degree and size of this method alone seems to indicate high method complexity and bad programming practices. \textit{Retrofit} is an HTTP client that builds a client service based on a Java interface augmented by HTTP-related annotations. The \textit{Builder\#parseParameterAnnotation} method matches the annotations and builds the appropriate \textit{ParameterHandler}. We recommend refactoring this method into smaller methods, with each method responsible for handling a particular annotation type and transforming it into a \textit{ParameterHandler}. This will improve code readability, reduce the method's complexity, and enhance the source code quality. It is worth noting that the outlier observed in the \textit{Builder\#parseParameterAnnotation} method could not be detected using the CCN model, as CCN operates at the abstract class level and does not consider method bodies or, in particular, local variables.

\begin{figure*}[h!]
    \centering
    \includegraphics[width=0.9\textwidth]{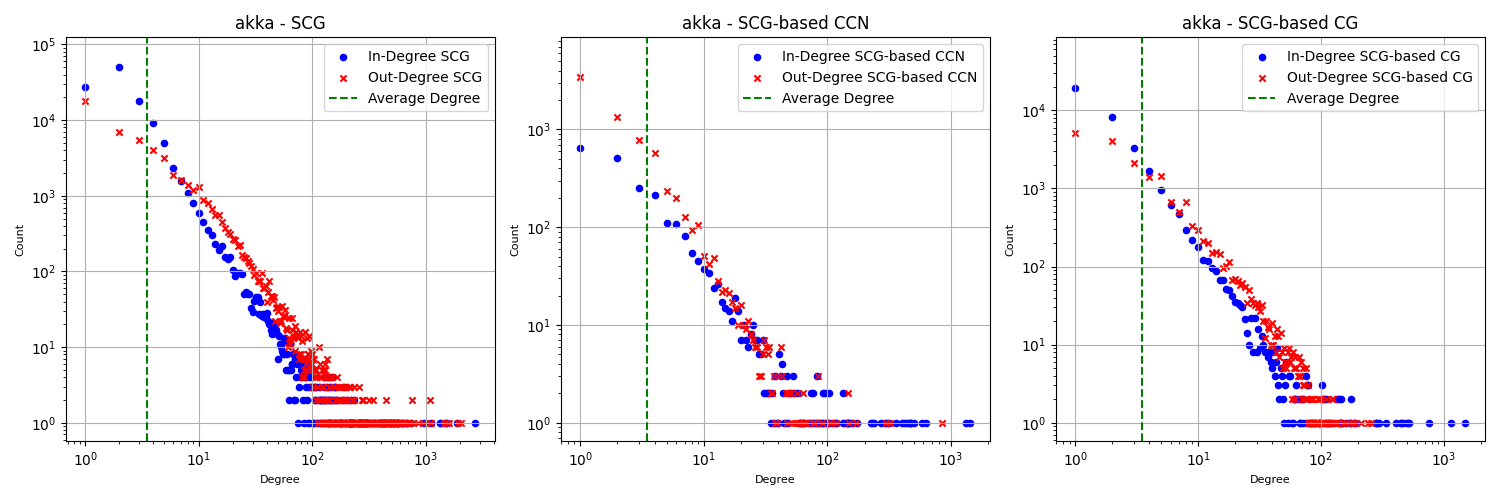}
    \caption{The incoming and outgoing degree distribution of nodes in the SCG, CCN, and CG graphs extracted for the \textit{akka} project. Nodes with high incoming degrees, such as the \textit{ActorRef} class and the method \textit{ActorRef\#!}, serve as the primary building blocks for the \textit{akka} project.}
    \label{fig:distribution-akka}
\end{figure*}

For the SCG graph, the highest Index of Dispersion can be observed for \textit{akka} project, with $IoD_{ID}=88.9$ and $IoD_{OD}=115.9$ which is much higher than in the case of all other projects. A~detailed distribution of node degrees, presented in Fig. \ref{fig:distribution-scg}, shows that a~significant number of nodes have degrees above 100 and there are several nodes with outgoing and incoming degrees above 1000. Notable nodes include the method responsible for sending messages in the \textit{akka} toolkit, \textit{ActorRef\#!} with an incoming degree of 2747, the \textit{ActorRef} class representing actor references with a degree of 1898, and the \textit{GraphStageLogic\#pull} method responsible for pulling new messages in \textit{akka} streams. Similar observations can be made for the CCN and CG networks. However, it is worth noting that CCN cannot identify critical methods, while CG falls short in identifying key class components. In this regard, the SCG information model provides the most comprehensive insight into important project outliers. Regarding $IoD_{OD}$, upon closer examination at node distribution presented at Fig. \ref{fig:distribution-akka}, it appears to be primarily influenced by large test classes. Test cases implemented with the \textit{scalatest} library reside within the test class body and contribute to the high outgoing node degree. Example include class \textit{ByteStringSpec} with 2031 outgoing connections and class \textit{ORMapSpec} with 1939 outgoing connections. We added also project characteristics computed for the \textit{akka} project without tests, which confirms big influence of test files on project structure and node distribution -- the value of $IoD_{OD}$ drops from 115.9 to 38.9. Interestingly, for CCN and CG models the network characteristics do not change much for \textit{akka} whether with or without tests. It may be an interesting avenue for future research to investigate the impact of tests on the project structure.

The low value of the Global Clustering Coefficient (GCC) indicates that the software forms a sparse network and does not manifest clustering tendency. A high GCC value would denote a high project clustering tendency, indicating that the network is well connected on the global level. On the other hand, the Average Clustering Coefficient (ACC) measures local clustering around individual nodes and places more weight on low-degree nodes. A high value of ACC would mean that the network is tightly connected, which violates good software practices and complicates software maintainability in the long run \cite{Bhattacharya2012}. All eleven analyzed projects have relatively low GCC and ACC values for both SCG and CG representations. Consequently, it should be expected that refactoring tasks related to software modularization should not be hindered by software structure. For CCN, the GCC values are higher, which is expected, as during CCN creation, the algorithm effectively applies path contraction to mark direct class dependencies and filter out entities other than classes. Class nodes contribute the most to the global structure and, hence, to clustering tendency. This naturally tightens the dependencies between entities and increases the clustering likelihood. 

\begin{figure}
  \centering

  \begin{subfigure}[h]{0.4\textwidth}
  \centering
  \includegraphics[width=200px]{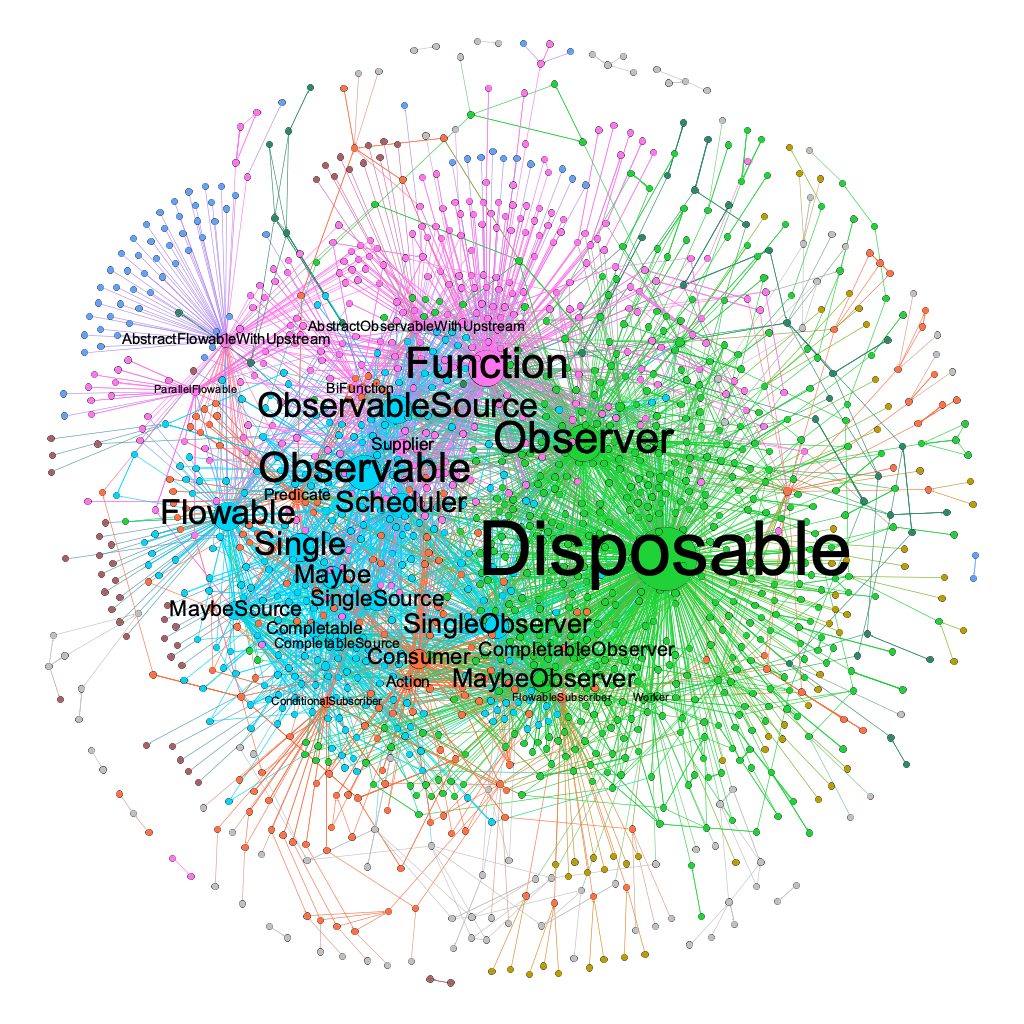}
    \caption{Visualization of the CCN for the \textit{RxJava} project reveals a clear tendency toward global clustering, confirmed by a high GCC value of 0.84. It also visually confirms the small-world network property. Low degree nodes on the outside contribute to an overall low ACC value.}
    \label{fig:RxJava-ccn}
  \end{subfigure}

  \begin{subfigure}[h]{0.4\textwidth}
  \centering
  \includegraphics[width=200px]{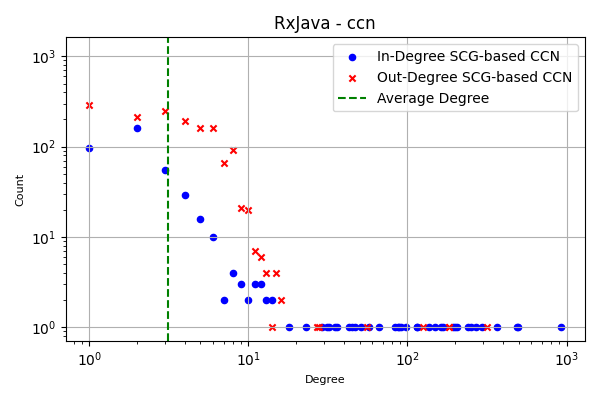}
    \caption{Degree Distribution for \textit{RxJava} project with visible high degree outliers.}
    \label{fig:distribution-RxJava-ccn}
  \end{subfigure}

  \caption{CCN visualization and degree distribution for the RxJava project confirming high tendency for global clustering.}
  \label{fig:RxJava-ccn-clustering}
\end{figure}

The RxJava project presents an interesting case, where we observe the largest spike of the GCC value (CCN model), reaching 0.84. High GCC and low ACC are typical characteristics of a small-world network, where most of the nodes are part of tightly-connected clusters, but there are a few hubs that bridge these clusters and shorten the path between otherwise distant nodes. A very high index of dispersion for incoming node degree equal to 299.7 might indicate that these hubs are nodes with particularly high incoming degree. Indeed, in Figure \ref{fig:RxJava-ccn-clustering}, we can visually observe the project's tendency for global clustering with a few hub nodes, such as \textit{Disposable}, \textit{Observer}, \textit{Function}, and \textit{ObservableSource}, being the top in-degree nodes. We can conclude that the CCN network, although extracted directly from the SCG model, provides an interesting perspective, showing the otherwise not visible clustering tendency of the \textit{RxJava} project at the class-level granularity.

These and similar metrics, apart from software comprehension activities, can also be used to assess the effort needed for certain refactoring activities or to measure the quality of the project structure throughout the evolution of the software. For example, increasing density and DAC with decreasing GCC can indicate project quality degradation.

\textit{\textbf{Conclusions.}} When comparing the SCG with the CG and CCN in the context of metrics-driven project structure comprehension, we can conclude that the perspectives provided by all three models can be useful, depending on the particular case. 
The CCN model is noteworthy for its emphasis on critical building blocks in object-oriented programming, unveiling compelling details such as the clustering tendency of the \textit{RxJava} class. Analysis of the CG graph, on the other hand, provides insights into the project structure from the perspective of project execution. However, being the superset of both CG and CCN, the SCG contains all these perspectives. Our experiments suggest that in order to gain a~comprehensive insight, it might be useful to compute various graph metrics for both the full SCG and the SCG-extracted CCN and CG graphs. 

\subsection{Discovering critical project entities}
\label{sec:project-critical-entities}

The most important nodes in the project are usually the starting points that developers look at to comprehend the project completely from the beginning. An ordered list of the most influential entities can help with a~more structured approach to reading the source code and learning about the project. Moreover, the most important nodes are those with the largest \textit{impact of change} \cite{Li2013}. In this case, the Semantic Code Graph representation, which is effectively the result of a~static structural dependence program analysis, can act as an intermediate program representation used in the Change Impact Analysis \cite{Lehnert2011}. Due to a potential high impact of change, when adjusting the source code of the top entities during software maintenance, the code review process should be conducted thoroughly with extra care to preserve high source code quality for the most critical parts of the system.  

We can assess the importance of project entities in a project by measuring the node size (in terms of the number of lines of code or by its outgoing/incoming degree), or by taking into account different metrics known in graph theory, such as graph centrality measures or node influence measures \cite{Gomez2019}. For all nodes in the SCG graphs of the studied projects, we have computed nine different properties that can serve as different metrics to determine the top nodes:

\newcounter{saveenum} 
\begin{enumerate}
    \item LOC -- top nodes by Lines-of-Code metrics. Code entities that span a significant number of lines of code naturally have a greater impact on the project.
    \item Outgoing Degree -- top nodes by the number of outgoing edges. Nodes with high degree suggest code entities aggregating big and likely important functionalities.
    \item Incoming Degree -- top nodes by the number of incoming edges. Many incoming edges denote extensive direct usage of this node throughout the project. 
    \item Eigenvector Centrality -- top nodes computed by the Eigenvector Centrality \cite{Bonacich1987} algorithm. High value means that a node is connected to multiple nodes that themselves have a high EC score. This metric focuses on direct connections.
    \item Katz Centrality -- top nodes computed by the Katz Centrality \cite{Katz1953} algorithm. High value of Katz Centrality metric is influenced not only by direct neighbors of a node but also by nodes  connected indirectly.
    \item Page Rank -- top nodes computed by Page Rank algorithm \cite{Brin1998}. Page Rank centrality is a variant of Eigenvector centrality which assigns a~PR value to a~node based on the number of nodes linking to this node and their PR values. This metric is affected by the importance of incoming links from highly ranked nodes.
    \item Betweenness Centrality -- top nodes computed by the Betweenness Centrality aglorithm \cite{Brandes2004}. High value of BC denotes that there are multiple shortest paths between other nodes going through this node. This metric indicates nodes crucial for the flow of information in the network.
    \item Harmonic Centrality -- top nodes computed by the Harmonic Centrality algorithm \cite{Rochat2009}  (Closeness Centrality metric for disconnected graphs). A node with high value of Harmonic Centrality is one that is close to many other nodes in the network. 
   \setcounter{saveenum}{\value{enumi}}
\end{enumerate}

Finally, we propose a combined importance metric that takes into account all metrics described above:
\begin{enumerate}
 \setcounter{enumi}{\value{saveenum}}
    \item Combined importance -- the metric is computed as follows: for each metric 1-8, take 20 nodes with the highest value (top 20 nodes). Assign scores to all positions using a reverse scoring system (position 1 = 20 points, position 20 = 1 point). Rank the nodes based on the sum of their scores. In other words, the top nodes will be the ones that scored highly in multiple metrics. 
\end{enumerate}

The most important nodes in the full SCG, Class Collaboration Network and the Call Graph in terms of different centrality metrics are presented in the Tables \ref{tab:scg-top-nodes-size-based}, \ref{tab:scg-top-nodes-influence-based}, and \ref{tab:scg-top-nodes-distance-base}. It should be noted that the resulting top nodes vary significantly for different metrics. In fact, metrics capture different aspects of potential node importance \cite{Gomez2019}. For example, high out-degree points to hubs, which can represent a structure encapsulating bigger chunk of related functionalities (such as class nodes), whereas high in-degree reveals nodes which are referenced by many other nodes and can be treated as project main building blocks. Metrics such as Eigenvector, Katz, or PageRank Centrality measure the importance of the node, which is influenced by the importance of its neighbors. Such nodes can be the most important code units from the maintenance point of view, where the introduced changes can have a great project-wide impact.  
Metrics such as Betweenness Centrality or Harmonic Centrality take into account the topological position in the graph and show the distance-wise central nodes through which many shortest paths lead, or nodes which are the closest nodes to multiple other nodes in the graph. These metrics might be especially meaningful for the call graph, indicating frequently called nodes on different execution paths.

\begin{table*}[h!]
\begin{scriptsize}
\setlength{\tabcolsep}{0.2em}
\renewcommand{\arraystretch}{1.1}

\centering
\sffamily
\fontsize{7pt}{6.5pt}\selectfont
\begin{tabular}{r|r|r|r|r|r|r|}

\cline{1-7}
 & 1. Lines of Code & \# & 2. Outgoing Degree & \# & 3. Incoming Degree & \# \\
\cline{1-7}
\multicolumn{7}{l|}{Semantic Code Graph}\\
\cdashline{1-7}
retrofit & RequestFactory & 800 & parseParameterAnnotation & 674 & Call & 147 \\ 
commons-io & IOUtils & 3608 & FileUtils & 176 & IOFileFilter & 131 \\ 
playframework & Multipart & 706 & PlayDocsValidation & 367 & Mapping & 882 \\ 
metals & MetalsLanguageServer & 2501 & updateWorkspaceDirectory & 503 & server & 157 \\ 
glide & BaseRequestOptions & 1398 & initializeDefaults & 261 & with & 183 \\ 
vert.x & HttpClientOptions & 1448 & MimeMapping & 990 & MimeMapping\#m & 989 \\ 
RxJava & Flowable & 20738 & Flowable & 541 & Disposable & 1251 \\ 
dubbo & URL & 1714 & PojoUtils\#realize1 & 390 & URL & 1891 \\ 
spring-boot & ServerProperties & 1757 & configureProperties & 300 & from & 330 \\ 
akka & Source & 3896 & ByteStringSpec & 2031 & ActorRef\#! & 2747 \\ 
spark & functions & 5380 & ScalaUDF & 2677 & Expression & 3561 \\ 

\cline{1-7}
\multicolumn{7}{l|}{Class Collaboration Network}\\
\cdashline{1-7}
retrofit & RequestFactory & 800 & HttpServiceMethod & 12 & Converter & 115 \\ 
commons-io & IOUtils & 3608 & FileFilterUtils & 48 & IOFileFilter & 102 \\ 
playframework & Multipart & 706 & Forms & 582 & Mapping & 876 \\ 
metals & MetalsLanguageServer & 2501 & MetalsLanguageServer & 29 & MetalsLanguageClient & 37 \\ 
glide & BaseRequestOptions & 1398 & Engine & 60 & Options & 143 \\ 
vert.x & HttpClientOptions & 1448 & FileSystemImpl & 110 & Handler & 615 \\ 
RxJava & Flowable & 20738 & Observable & 317 & Disposable & 927 \\ 
dubbo & URL & 1714 & RegistryProtocol & 90 & URL & 1250 \\ 
spring-boot & ServerProperties & 1757 & Binder & 78 & ConfigurationPropertyName & 150 \\ 
akka & Source & 3896 & GraphApply & 853 & ActorRef & 1441 \\ 
spark & functions & 5380 & functions & 851 & Expression & 2922 \\ 

\cline{1-7}
\multicolumn{7}{l|}{Call Graph}\\
\cdashline{1-7}
retrofit & parseParameterAnnotation & 456 & parseParameterAnnotation & 417 & Builder\#method & 77 \\ 
commons-io & FilenameUtils\#doNormalize & 96 & FilenameUtils\#doNormalize & 119 & CloseableURLConnection\#urlConnection & 46 \\
playframework & webSocketProtocol & 349 & forwardsRouter\#apply & 120 & unbindAndValidate & 257 \\ 
metals & updateWorkspaceDirectory & 419 & updateWorkspaceDirectory & 431 & XtensionJavaFuture\#asScala & 101 \\ 
glide & initializeDefaults & 245 & initializeDefaults & 183 & Glide\#with & 182 \\ 
vert.x & EventBusOptionsConverter\#fromJson & 254 & TCPServerBase\#listen & 170 & Future\#onComplete & 170 \\  
RxJava & MemoryPerf\#main & 378 & ConcatMapEagerMainObserver\#drain & 134 & Exceptions\#throwIfFatal & 347 \\ 
dubbo & PojoUtils\#realize1 & 249 & PojoUtils\#realize1 & 229 & StringUtils\#isEmpty & 418 \\
spring-boot & MavenBuild\#execute & 73 & configureProperties & 233 & PropertyMapper\#from & 329 \\ 
akka & UnzipWith22\#createLogic & 476 & DistributedPubSubMediator\#receive & 256 & GraphStageLogic\#pull & 1501 \\ 
spark & prepareSubmitEnvironment & 641 & prepareSubmitEnvironment & 679 & Params\#\$ & 1382 \\ 

\end{tabular}

\end{scriptsize}
\caption{Top nodes in the studied projects in terms of general properties (1. Lines of Code, 2. Outgoing Degree, 3. Incoming Degree);}
\label{tab:scg-top-nodes-size-based}
\end{table*}

\begin{table*}[!htb]
\begin{scriptsize}
\setlength{\tabcolsep}{0.2em}
\renewcommand{\arraystretch}{1.1}

\centering
\sffamily
\fontsize{7pt}{6.5pt}\selectfont
\begin{tabular}{r|r|r|r|r|r|r|}

\cline{1-7}
 & 4. Eigenvector Centrality & \# & 5. Katz Centrality & \# & 6. Page Rank & \# \\
\cline{1-7}
\multicolumn{7}{l|}{Semantic Code Graph} \\
\cdashline{1-7}
retrofit & Utils\#toResolve & 0.6051 & Call & 2.5066 & Call & 0.0180 \\ 
commons-io & Tailer\#tailable & 0.2603 & IOFileFilter & 2.3425 & IOFileFilter & 0.0131 \\ 
playframework & Request\#[A] & 0.2977 & Mapping & 10.1967 & Mapping & 0.0105 \\ 
metals & Version\#major. & 0.4237 & BaseLspSuite\#server & 2.5840 & AnsiStateMachine\#apply & 0.0038 \\ 
glide & TranscodeType & 0.3777 & Glide\#with & 2.8629 & Key & 0.0069 \\ 
vert.x & HttpMethod & 0.4860 & MimeMapping\#m & 10.8902 & HttpMethod & 0.0143 \\ 
RxJava & Flowable & 0.5522 & Disposable & 13.9548 & Disposable & 0.0370 \\ 
dubbo & URL\#urlAddress & 0.3399 & URL & 20.9279 & URL & 0.0201 \\ 
spring-boot & Regex\#group & 0.5066 & PropertyMapper\#from & 4.3904 & DependencyVersion & 0.0050 \\ 
akka & Source & 0.4295 & ActorRef\#! & 29.0573 & ReplicatedData & 0.0246 \\ 
spark & ConfigBuilder\#version & 0.3444 & Expression & 37.9700 & Expression & 0.0158 \\

\cline{1-7}
\multicolumn{7}{l|}{Class Collaboration Network}\\
\cdashline{1-7}
retrofit & Converter & 0.8521 & Converter & 2.1933 & Callback & 0.2146 \\ 
commons-io & TailerListener & 0.8256 & IOFileFilter & 2.0228 & PathFilter & 0.0538 \\ 
playframework & Binding & 0.9392 & Mapping & 9.7610 & TypedEntry & 0.3800 \\ 
metals & Cancelable & 0.5952 & MetalsLanguageClient & 1.3770 & Cancelable & 0.0174 \\ 
glide & Transformation & 0.3522 & Options & 2.4766 & Editor & 0.0605 \\ 
vert.x & Handler & 0.7364 & Handler & 8.3021 & Buffer & 0.1275 \\ 
RxJava & Scheduler & 0.6571 & Disposable & 10.7265 & ParallelFlowable & 0.1242 \\ 
dubbo & ApplicationModel & 0.4460 & URL & 13.9248 & URLAddress & 0.2983 \\ 
spring-boot & ConditionMessage & 0.7171 & ConfigurationPropertyName & 2.5881 & Builder & 0.2276 \\ 
akka & ForwardOps & 0.4815 & Graph & 16.1818 & ActorRef & 0.1371 \\ 
spark & Column & 0.5824 & Expression & 31.2843 & Param & 0.1191 \\ 

\cline{1-7}
\multicolumn{7}{l|}{Call Graph}\\
\cdashline{1-7}
retrofit & Utils\#resolve & 0.4450 & Builder\#method & 1.7843 & Utils\#getRawType & 0.0124 \\ 
commons-io & FileAlterationObserver\#listFiles & 0.8323 & CloseableURLConnection\#urlConnection & 1.4605 & AbstractFileFilter\#accept & 0.0159 \\ 
playframework & Multipart\#partStart & 0.3389 & Mapping\#bind & 3.5704 & Configuration\#underlying. & 0.0018 \\ 
metals & SemanticdbTreePrinter\#printSymbol & 0.4449 & XtensionJavaFuture\#asScala & 2.0247 & XtensionAbsolutePath\#path & 0.0028 \\ 
glide & RequestBuilder\#thumbnailBuilder & 0.4773 & Glide\#with & 2.8308 & Preconditions\#checkNotNull & 0.0040 \\ 
vert.x & JsonParserImpl\#handleEvent & 0.4855 & Future\#onComplete & 2.7083 & JsonObject\#map & 0.0065 \\ 
RxJava & MergeObserver\#observers & 0.5741 & Exceptions\#throwIfFatal & 4.4854 & DISPOSED & 0.0040 \\ 
dubbo & JavaBeanDescriptor & 0.5246 & StringUtils\#isEmpty & 5.3694 & URL\#getParameter & 0.0068 \\ 
spring-boot & JSONTokener\#pos & 0.7939 & PropertyMapper\#from & 4.3362 & PropertiesConfigAdapter\#properties& 0.0012 \\ 
akka & WorkPullingProducerControllerImpl\#context. & 0.3596 & GraphStageLogic\#pull & 16.4494 & ActorRef\#! & 0.0021 \\ 
spark & StructField\#name & 0.8183 & Params\#\$ & 15.0390 & AnalysisException\#<init> & 0.0030 \\

\end{tabular}

\end{scriptsize}
\caption{Top nodes in the studied projects in terms of influence-based centrality metrics (4. Eigenvector Centrality, 5. Katz Centrality, 6. Page Rank);}
\label{tab:scg-top-nodes-influence-based}
\end{table*}

\begin{table*}[!htb]
\begin{scriptsize}
\setlength{\tabcolsep}{0.2em}
\renewcommand{\arraystretch}{1.1}

\centering
\sffamily
\fontsize{6.5pt}{6.5pt}\selectfont
\begin{tabular}{r|r|r|r|r|r|r|}

\cline{1-7}
 & 7. Betweenness Centrality* & \# & 8. Harmonic Centrality & \# & \textbf{9. Combined importance} & \# \\
\cline{1-7}
\multicolumn{7}{l|}{Semantic Code Graph} \\
\cdashline{1-7}
retrofit & Retrofit & 537k & Builder\#parseParameterAnnotation & 0.1023 & Retrofit & 5 \\ 
commons-io & IOStreams\#forAll & 301k & FileUtils & 0.0924 & IOConsumer & 4 \\ 
playframework & RequestHeader & 3 434k & Forms & 0.0622 & RequestHeader & 5 \\ 
metals & Ammonite & 6 789k & MetalsLanguageServer & 0.1951 & MetalsLanguageServer & 4 \\ 
glide & Glide & 7 762k & Glide & 0.0854 & RequestBuilder & 6 \\ 
vert.x & VertxImpl & 45 953k & VertxImpl & 0.1208 & VertxImpl & 4 \\ 
RxJava & Flowable & 18 802k & Flowable & 0.0673 & Flowable & 7 \\ 
dubbo & URL & 79 424k & DubboBootstrap & 0.0450 & URL & 7 \\ 
spring-boot & Binder & 1 107k & BuildImageMojo\#buildImage & 0.0106 & ConfigurationPropertyName & 5 \\ 
akka & TraversalBuilder & 106 628k & SystemMessageDeliverySpec & 0.0526 & ActorRef & 3 \\ 
spark & SparkContext & 714 504k & BaseSessionStateBuilder & 0.1034 & Expression & 4 \\ 

\cline{1-7}
\multicolumn{7}{l|}{Class Collaboration Network}\\
\cdashline{1-7}
retrofit & Retrofit & 0.243k & HttpServiceMethod & 0.0490 & Retrofit & 8 \\ 
commons-io & AbstractFileFilter & 0.075k & Uncheck & 0.0636 & IOFileFilter & 4 \\ 
playframework & Result & 2.4k & Helpers & 0.0377 & RequestHeader & 5 \\ 
metals & BuildTargets & 1.2k & DebugProvider & 0.0773 & BuildTargets & 4 \\ 
glide & Glide & 6.0k & Glide & 0.0827 & RequestManager & 5 \\ 
vert.x & Vertx & 13.7k & VertxImpl & 0.1507 & JsonObject & 4 \\ 
RxJava & Flowable & 12.5k & Single & 0.0197 & Flowable & 8 \\ 
dubbo & ApplicationModel & 51.0k & DubboBootstrap & 0.0355 & ApplicationModel & 5 \\ 
spring-boot & SpringApplication & 1.2k & Builder & 0.0095 & ConfigurationPropertyName & 4 \\ 
akka & Props & 2.7k & UnzipWithApply & 0.0084 & ActorRef & 4 \\ 
spark & SparkContext & 256.1k & SparkSession & 0.0200 & Expression & 5 \\

\cline{1-7}
\multicolumn{7}{l|}{Call Graph}\\
\cdashline{1-7}
retrofit & Retrofit\#create & 15k & Builder\#parseParameterAnnotation & 0.0895 & Builder\#parseParameterAnnotation & 4 \\ 
commons-io & FilenameUtils\#getPrefixLength & 2k & XmlStreamReader & 0.0131 & FilenameUtils\#doNormalize & 4 \\ 
playframework & RangeResult\#ofSource & 1k & PlayRequestHandler\#handle & 0.0127 & Configuration\#get & 3 \\ 
metals & MetalsLanguageServer\#executeCommand & 35k & MetalsLanguageServer\#executeCommand & 0.0627 & MetalsLanguageServer\#executeCommand & 4 \\ 
glide & Glide\#initializeGlide & 107k & MultiRequestTest\#(...)whenRequestListenerIsCalled & 0.0167 & Glide\#with & 4 \\ 
vert.x & TCPSSLOptions & 16k & BareCommand\#startVertx & 0.0108 & ContextInternal\#promise & 3 \\ 
RxJava & SpscLinkedArrayQueue & 1690 & MemoryPerf\#main & 0.0028 & DisposableHelper\#dispose & 4 \\ 
dubbo & FrameworkModel\#defaultModel & 1 408k & ReferenceConfig\#init & 0.0113 & Node\#getUrl & 3 \\ 
spring-boot & SpringApplication\#run & 30k & BuildImageMojo\#buildImage & 0.0061 & SpringApplication\#run & 4 \\ 
akka & GraphStageLogic\#internalCompleteStage & 137k & ClusterCoreDaemon\#initialized() & 0.0034 & GraphStageLogic\#pull & 4 \\ 
spark & SparkContext\#clean & 242k & ALSExample\#main & 0.0052 & Params\#set & 3 \\ 

\end{tabular}

\end{scriptsize}
\caption{Top nodes in the studied projects in terms of distance-based centrality metrics (7. Betweennness Centrality, 8. Harmonic Centrality) and according to the proposed 9. \textit{combined metric}. 
*the value of Betweennes Centrality was not normalized and the total number of shortest paths going through a~particular node is used for more convenient result presentation.}
\label{tab:scg-top-nodes-distance-base}
\end{table*}

Finally, there are nodes ranked highly according to multiple metrics, indicating units particularly important for the entire project. Our \textit{combined importance} metric was proposed with the purpose of finding such cases.  
The top three nodes according to the \textit{combined importance} metric for each project are presented in Tab. \ref{fig:scg-top-nodes-combined}. It is particularly interesting to observe the difference in results obtained from the Full SCG versus the Call Graph and Class Collaboration Network. For example, the critical SCG nodes for the \textit{metals} project are classes \textit{MetalsLanguageServer}, \textit{BuildTargets}, and \textit{MetalsLanguageClient}. These nodes are important from the \textit{project structure perspective}, i.e., for the purpose of software comprehension in the context of project understanding or planning refactoring activities. Indeed, \textit{metals} is a Language Server Protocol implementation for Scala, where \textit{MetalsLanguageServer\#languageClient} represents the tool (IDE) and \textit{MetalsLanguageServer} implements all the functionalities required by the client, so these are two most important and complementary classes. \textit{BuildTargets} is a non-obvious candidate for the important node, but on the closer inspection we can learn that \textit{BuildTargets} is a in-memory cache for looking up build server metadata and is involved in most of the \textit{metals} functionalities. For the Class Collaboration Network, \textit{BuildTargets} is also the most important node, followed by \textit{Cancelable} and \textit{Compilers}. \textit{Cancelable} is a trait that is extended by more than twenty different classes, making it an important node from the class inheritance perspective. The \textit{Compilers} class is used in eight different places as a constructor argument for eight top-level classes in \textit{Metals}, and because of that, it has a high influence score (it is a top node according to Eigenvector Centrality and Page Rank metrics).
On the other hand, the results obtained from the Call Graph clearly reflect \textit{the runtime perspective}, focusing on invocation dependencies. For the \textit{metals} project, the important nodes in the Call Graph are the \textit{MetalsLanguageServer\#executeCommand} method which is responsible for executing a~large set of commands, the \textit{MetalsLanguageServer\#updateWorkspaceDirectory} method responsible for initialization of almost forty different providers in \textit{MetalsLanguageServer}, and the \textit{XtensionJavaFuture\#asScala} helper method used in all the places when \textit{metals} interacts directly with the underlying Java based LSP4J\footnote{\url{https://projects.eclipse.org/projects/technology.lsp4j}} client. These methods are critical from the runtime, and hence performance and program execution correctness, perspective. 

\begin{table*}[!htb]
\begin{footnotesize}
\setlength{\tabcolsep}{0.2em}
\renewcommand{\arraystretch}{1.1}

\centering
\sffamily
\fontsize{7pt}{6.5pt}\selectfont
\begin{tabular}{r|r|r|r|r|r|r|}

\cline{1-7}
Name & \#1 & \# & \#2 & \# & \#3 & \# \\
\cline{1-7}
\multicolumn{7}{l|}{Semantic Code Graph} \\
\cdashline{1-7}
retrofit & Retrofit & 5 & parseParameterAnnotation & 4 & Converter & 4 \\ 
commons-io & IOConsumer & 4 & IOCase & 3 & IOStream & 3 \\ 
playframework & RequestHeader & 5 & Result & 4 & AkkaHttpServer & 3 \\ 
metals & MetalsLanguageServer & 4 & BuildTargets & 2 & MetalsLanguageServer\#languageClient. & 2 \\ 
glide & RequestBuilder & 7 & RequestManager & 4 & Request & 3 \\ 
vert.x & VertxImpl & 4 & ContextInternal & 4 & Future & 4 \\  
RxJava & Flowable & 7 & Observable & 6 & Single & 5 \\ 
dubbo & URL & 7 & ApplicationModel & 4 & Invoker & 3 \\ 
spring-boot & ConfigurationPropertyName & 5 & SpringApplication & 3 & ConfigurationPropertySource & 3 \\ 
akka & ActorRef & 3 & Source & 3 & ActorRef\#! & 2 \\
spark & Expression & 4 & SparkContext & 3 & RDD & 3 \\ 

\cline{1-7}
\multicolumn{7}{l|}{Class Collaboration Network}\\
\cdashline{1-7}
retrofit & Retrofit & 8 & Factory & 6 & Call & 5 \\ 
commons-io & IOFileFilter & 4 & IOStream & 4 & AbstractFileFilter & 4 \\ 
playframework & RequestHeader & 5 & Result & 5 & PlayBodyParsers & 4 \\ 
metals & BuildTargets & 4 & Cancelable & 4 & Compilers & 4 \\ 
glide & RequestManager & 5 & RequestBuilder & 5 & Key & 4 \\ 
vert.x & JsonObject & 4 & Vertx & 4 & Handler & 4 \\ 
RxJava & Flowable & 8 & Single & 6 & Maybe & 5 \\ 
dubbo & ApplicationModel & 5 & URL & 5 & ModuleModel & 5 \\ 
spring-boot & ConfigurationPropertyName & 4 & ServerProperties & 4 & ConditionOutcome & 4 \\ 
akka & ActorRef & 4 & SubFlow & 3 & GraphStageLogic & 3 \\ 
spark & Expression & 5 & DataType & 4 & RDD & 4 \\

\cline{1-7}
\multicolumn{7}{l|}{Call Graph}\\
\cdashline{1-7}
retrofit & parseParameterAnnotation & 4 & Utils\#resolve & 4 & RequestFactory.Builder\#build & 4 \\ 
commons-io & FilenameUtils\#doNormalize & 4 & XmlStreamReader\#calculateHttpEncoding & 3 & IOBaseStream\#unwrap & 3 \\ 
playframework & Configuration\#get & 3 & RequestHeader\#attrs & 3 & Route\#call & 3 \\
metals & MetalsLanguageServer\#executeCommand & 4 & updateWorkspaceDirectory & 3 & XtensionJavaFuture\#asScala & 3 \\ 
glide & Glide\#with & 4 & Preconditions\#checkNotNull & 3 & LoadBytesTest\#context & 3 \\ 
vert.x & TCPServerBase\#listen & 3 & ContextInternal\#promise & 3 & BufferImpl\#buffer & 3 \\ 
RxJava & DisposableHelper\#dispose & 4 & ConcatMapEagerMainObserver\#drain & 4 & DisposableHelper\#setOnce & 4 \\ 
dubbo & Node\#getUrl & 3 & DubboCodec\#decodeBody & 3 & StringUtils\#isEmpty & 3 \\ 
spring-boot & SpringApplication\#run & 4 & PropertyMapper\#get & 3 & FlywayConfiguration\#configureProperties & 3 \\ 
akka & GraphStageLogic\#pull & 4 & GraphStageLogic\#completeStage & 4 & ActorRef\#! & 3 \\ 
spark & Column\#expr & 3 & Params\#set & 3 & Expression\#dataType & 3 \\ 

\end{tabular}

\end{footnotesize}
\caption{Top 3 nodes in the studied projects according to the \textbf{combined importance} calculated for the full SCG, the Class Collaboration Network, and the Call Graph.}
\label{fig:scg-top-nodes-combined}
\end{table*}

\begin{figure}[!htb]
    \centering
    \includegraphics[width=0.45\textwidth]{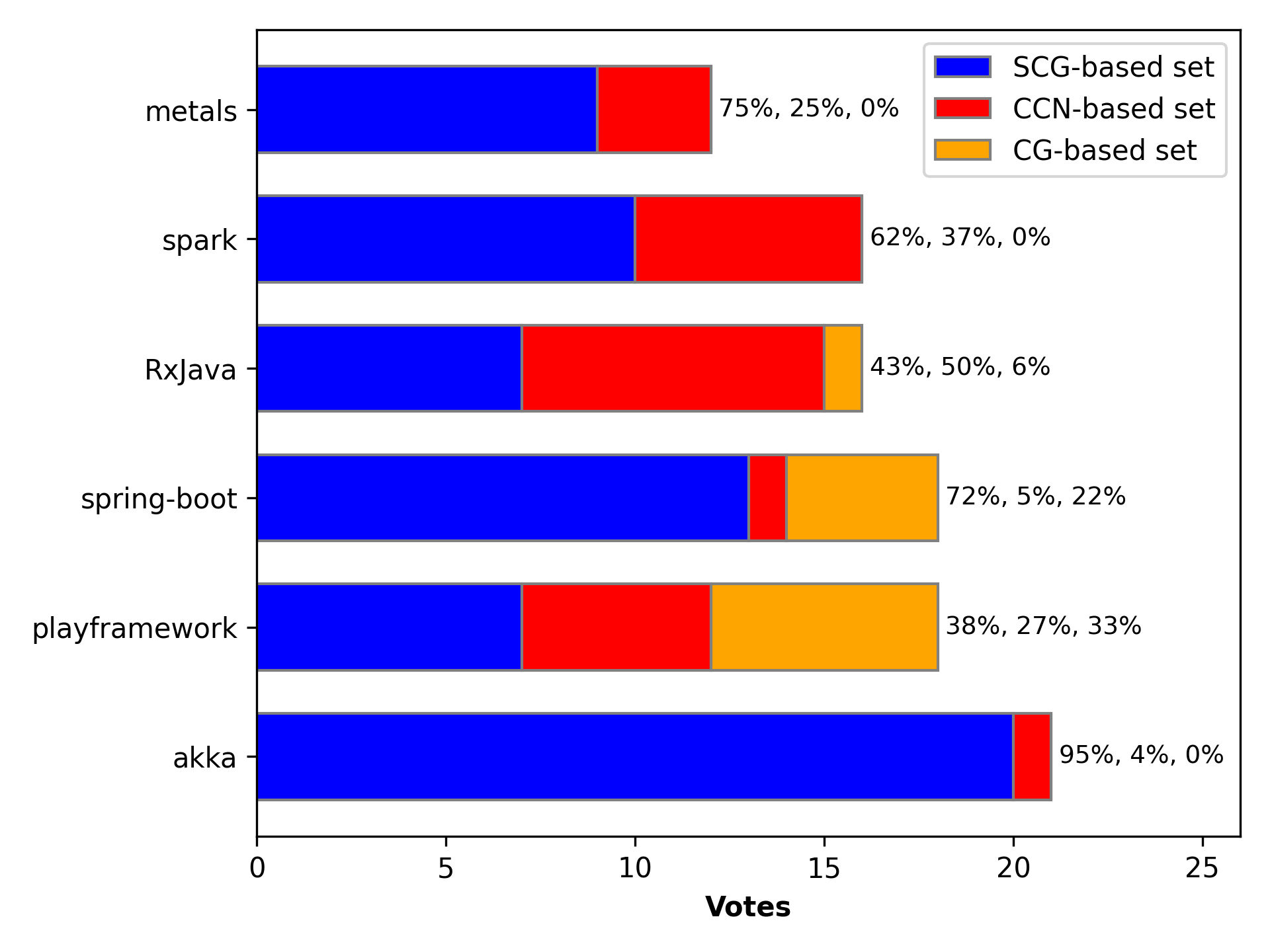}
    \caption{The survey results to evaluate the performance of SCG, CCN and CG in finding crucial entities. Participants were asked the question, "What is the most important entity set in terms of software maintenance for a given project?" For each project, participants were provided with three sets of the three most important nodes computed using the SCG, CCN, and CG models, respectively. Results are presented only for projects with more than 10 answers.}
    \label{fig:crucial-nodes-survey-results}
\end{figure}

To evaluate the performance of the SCG model in finding critical entities compared to CCN and CG, we conducted a survey based on the data presented in Table \ref{fig:scg-top-nodes-combined}. The survey posed a single question: "What is the most important entity set in terms of software maintenance for a given project?" For each of the eleven projects, participants were presented with three options to choose from, i.e., the top three combined nodes for SCG, CCN, or CG, respectively. Participants were instructed to vote only if they were familiar with the project in question. The results are presented in Fig. \ref{fig:crucial-nodes-survey-results}.



We received a total of 26 survey responses; however, some projects garnered fewer answers, likely due to lesser familiarity among the respondents. 
For further quantitative analysis we considered only projects with 10 or more responses, thus excluding \textit{dubbo} (two answers), \textit{glide} (two answers), \textit{commons-io} (six answers), \textit{retrofit} (six answers), and \textit{vert.x} (eight answers) projects. 

\begin{figure}[!htb]
    \centering
    \includegraphics[width=0.45\textwidth]{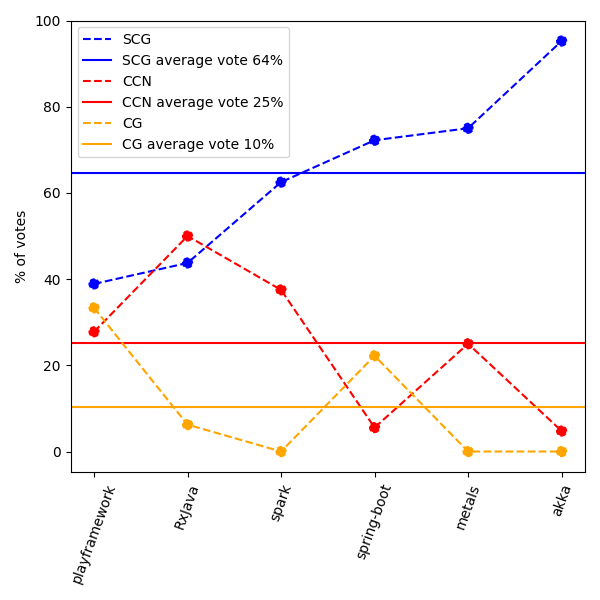}
    \caption{The comparison of SCG, CCN and CG performance for crucial nodes discovery based on survey results presented in Fig. \ref{fig:crucial-nodes-survey-results}. The figure illustrates the percentage of votes assigned to each set of nodes for every project with 10 or more responses in the survey. This provides a detailed overview of the user preferences and highlights the comparative performance of SCG, CCN, and CG in terms of crucial nodes discovery.}
    \label{fig:crucial-nodes-survey-comparison}
\end{figure}

\emph{\textbf{Conclusions.}} An overview of survey results is presented in Fig. \ref{fig:crucial-nodes-survey-comparison}. On average, users pointed to SCG-based results as the most important set 64\% of the time, while CCN and CG models were chosen 25\% and 10\% of the time, respectively. Interestingly, for every project except \textit{RxJava}, the SCG-based set received the highest number of votes. Consequently, the survey indicates that the SCG model led to more accurate results in comparison with CCN and CG models. 

\subsection{Interactive visualizations and answering reachability questions}
\label{sec:reachability}

\begin{figure*}[!htb]
\centering
\includegraphics[width=500px]{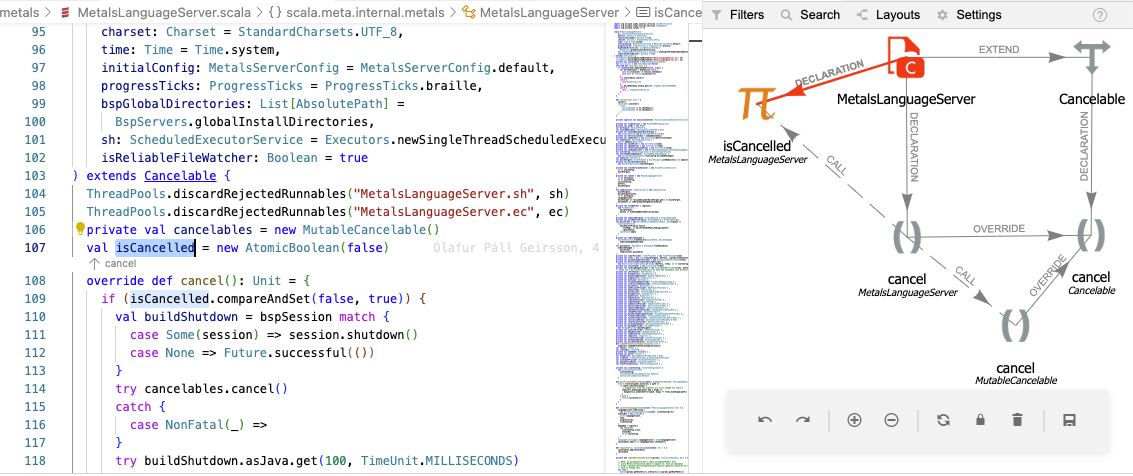}
\caption{Visual Studio Code IDE with Graph Buddy, a tool for interactive source code visualization based on the SCG information model. The screenshot shows highlighted \textit{isCancelled} \textit{declaration} edge in the graph and the corresponding code.}
\label{fig:gb-scg-rich-model}
\end{figure*}

SCG can be used as a base model to visualize the dependencies of the source code entities and answer different reachability questions \cite{LaToza2010} effectively. Thanks to SCG, showing a~call hierarchy is simply a matter of traversing the \textit{CALL} edges from the given starting point node. Finding a path between two nodes can be solved with any shortest path algorithm. Due to SCG richness,  interactive browsing can be augmented with structural details to show where the nodes are defined or what other important code relations are involved. Also, with SCG, we can create documentation of parts of the system and share it with other developers. These functionalities can be implemented in close relation to the source code thanks to the required SCG \textit{location} property of each node and edge. The SCG was utilized in this way in Graph Buddy \footnote{\url{https://github.com/VirtusLab/graphbuddy}}, a~code browsing and visualization tool (Figure \ref{fig:gb-scg-rich-model}). Answering reachability questions and documenting parts of the system with Graph Buddy was described in detail in \cite{Borowski2022}. 

\begin{figure}[!h]
\centering
\begin{subfigure}[h]{0.4\textwidth}
  \centering
  \begin{lstlisting}[language=Scala, basicstyle=\footnotesize]
// ScalaGreeting.scala
class ScalaGreeting() {
  def greeting(): String = 
    "Scala Greeting"
}
// JavaGreeting.java
public class JavaGreeting {
    public String greeting() {
        return "Java Greeting!";
    }
}
// Main.scala
object Hello extends App {
  val scalaGreeting = 
    new ScalaGreeting()
  val javaGreeting = 
    new JavaGreeting()
  println(
    scalaGreeting.greeting()
  )
  println(
    javaGreeting.greeting()
  )
}
  \end{lstlisting}
  \caption{Scala and Java code mixed in one project.}
  \label{fig:java-and-scala}
\end{subfigure}%

\begin{subfigure}[h]{0.4\textwidth}
  \centering
  \includegraphics[width=200px]{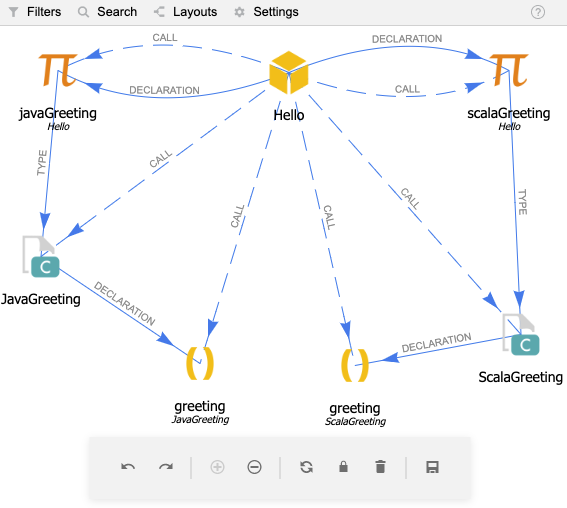}
    \caption{Interactive visualization of multi-language Semantic Code Graph in Graph Buddy tool.}
    \label{fig:gb-visualization-interop}
\end{subfigure}
\caption{The Semantic Code Graph generates stable element identifiers that allows for multilanguage analysis and visualization.}
\label{fig:scala-java-graphbuddy}
\end{figure}

Using Graph Buddy, developers could complete typical code browsing tasks in a more convenient way than using features available in popular IDEs. Moreover, thanks to language-independence of the SCG model and the algorithm for stable identifiers described in section \ref{scg-stable-system-identifiers}, this tool uniquely supports multilingual projects that contain Java and Scala source code. Fig. \ref{fig:scala-java-graphbuddy} presents the visualization of our demo project\footnote{\url{https://github.com/liosedhel/semantic-code-graph-jvm-interop}} where the Java class \textit{JavaGreeting} is instantiated and later called from the Scala \textit{Hello} companion object.

\textit{\textbf{Conclusions.}} SCG-powered interactive visualization was described in detail in our previous work \cite{Borowski2022}, where we also presented its quantitative evaluation. To this end, we asked ten programmers to complete three programming tasks using Graph Buddy, all involving the metals project : (1) finding a specific call hierarchy, (2) finding a call path, (3) learning about a~specific code entity. The participants achieved a very high success rate of 80-100\% and in a~survey they agreed that at least two of the tasks would be difficult to solve without Graph Buddy. 
Similar interactive browsing capabilities would not be possible to achieve with either CCN or CG models, as they represent only a portion of code entities and they lack a~direct connection to the source code. 

\subsection{Finding code similarities}
\label{sec:finding-similarities}

Finding code similarities in the project can help both in software comprehension and directly in software maintenance. To better understand the software, we can look for cases of similar code in the project in order to discover not obvious relations and analyze subtle differences between them \cite{Middleton2022}. In software maintenance, we use clone detection to find and refactor code duplication \cite{fowler1997refactoring} or to apply bug fixes in multiple places.
It has been proven that graph based representation of the source code (such as Program Dependence Graph \cite{Ferrante1987}) can be used to find code similarities \cite{Krinke2001, Kim2016}. Using the SCG graph, we can find \textit{semantically similar} code fragments within the project. Existing algorithms are typically based on computing graph similarity \cite{Krinke2001, roy2007survey, Kim2016}. However, the disadvantage of this approach is its high computational complexity and, consequently, slow performance for large code bases. We propose a~simple and effective heuristic for quick finding similarity between methods in a~given project using the SCG.  The heuristic, fully presented in Listing \ref{lst:similarity-algorithm}, can be summarized as follows:
\begin{itemize}
    \item \textbf{Step 1.} Generate a list of all method pairs in the project.
    \item \textbf{Step 2.} For each method in a pair, retrieve a set of its adjacent nodes; then compute the intersection of the sets.
    \item \textbf{Step 3.} Determine the similarity between methods based on the size of their intersection set. Two nodes are treated as similar if for both nodes the minimal size of the intersection is higher than  \texttt{s\_min} and the fraction of the intersecting nodes (within all adjacent nodes) is higher than \texttt{s\_min\_p}.
    \item \textbf{Step 4.} Discard similar methods having the same parent to exclude similarities occurring in the same code unit and focus on non obvious ones. 
\end{itemize}

\begin{lstlisting}[basicstyle=\footnotesize, language=Python, caption=Algorithm for finding similar methods in a project., label={lst:similarity-algorithm}]
import itertools as it
import scg

# min. 5 nodes in intersection
s_min = 5
# min. 50% of method nodes has to be the same
s_min_p = 50
scg_files = scg.read_scg("data/metals")
G = scg.create_graph(scg_files)
# find all method nodes

# Step 1.
methods = [
    n
    for n in G.nodes()
    if G.nodes[n].get("scg_node").kind == "METHOD"
]
# generate all method pairs
methods_comb = list(it.combinations(methods, 2))

# find defining node for given node n
def parent(G, n):
    return [
        p
        for p in G.predecessors(n)
        if G[p][n]["type"] == "DECLARATION"
    ]


similar = []
for (
    n1,
    n2,
) in methods_comb:
    # Step 2.
    n1_n = set(G.successors(n1))
    n2_n = set(G.successors(n2))
    s = len(set(n1_n).intersection(n2_n))
    # fraction of common nodes in each method
    n1_p = 0 if s==0 else int(s / len(n1_n) * 100)
    n2_p = 0 if s==0 else int(s / len(n2_n) * 100)
    # Step 3.
    if s >= s_min and (
        n1_p >= s_min_p and n2_p >= s_min_p
    ):
        # Step 4.
        if parent(G, n1) != parent(G, n2):
            similar.append((n1,n2,s,n1_s,n2_s))
# 'similar' contains pairs of similar methods            
print(similar)
\end{lstlisting}

Even such a~simplified heuristic can reveal interesting similarities, for example in the \textit{metals} project, as shown in Fig. \ref{fig:metals-similar-methods}. Two similar methods --\textit{oldReloadResult} and \textit{oldInstallResult} -- were found which could be refactored into  a~\textit{common} method presented in Listing \ref{lst:common}.

\begin{figure}[!htb]
\begin{lstlisting}[basicstyle=\footnotesize, language=Scala, caption=\textit{metals} method from \textit{WorkspaceReload} class., label={lst:oldReloadResult}]
def oldReloadResult(
    digest: String
): Option[WorkspaceLoadedStatus] = {
    if (tables.dismissedNotifications
            .ImportChanges.isDismissed) {
      Some(WorkspaceLoadedStatus.Dismissed)
    } else {
      tables.digests.last().collect {
        case Digest(md5, status, _) 
            if md5 == digest =>
          WorkspaceLoadedStatus.Duplicate(status)
      }
    }
}
\end{lstlisting}
\begin{lstlisting}[basicstyle=\footnotesize, language=Scala, caption=\textit{metals} method from \textit{BloopInstall} class., label={lst:oldInstallResult}]
private def oldInstallResult(
  digest: String
): Option[WorkspaceLoadedStatus] = {
    if (notification.isDismissed) {
      Some(WorkspaceLoadedStatus.Dismissed)
    } else {
      tables.digests.last().collect {
        case Digest(md5, status, _) 
            if md5 == digest =>
          WorkspaceLoadedStatus.Duplicate(status)
      }
    }
}
\end{lstlisting}
\begin{lstlisting}[basicstyle=\footnotesize, language=Scala, caption=Extracted common code from \textit{oldReloadResult} and \textit{oldInstallResult} methods., label={lst:common}]
private def common(
  digest: String,
  condition: Boolean,
  tables: Tables
): Option[WorkspaceLoadedStatus] = {
    if (condition) {
      Some(WorkspaceLoadedStatus.Dismissed)
    } else {
      tables.digests.last().collect {
        case Digest(md5, status, _) 
            if md5 == digest =>
          WorkspaceLoadedStatus.Duplicate(status)
      }
    }
}
\end{lstlisting}
    \caption{Method \textit{oldReloadResult} and \textit{oldReloadResult} from \textit{metals} project are similar, differing only in naming and the first condition statement. Consequently, a~\textit{common} method shown in Listing \ref{lst:common} can be used to replace them.}
    \label{fig:metals-similar-methods}
\end{figure}

\emph{\textbf{Conclusions.}} Listing \ref{lst:similarity-algorithm} demonstrates how code analysis algorithms can be implemented using the raw SCG graph data structures. The example reveals code similarity computed based on the code structure represented in the SCG model, containing entities and relations not available in CCN or CG alone (e.g., \textit{PARAMETER}, \textit{RETURN\_TYPE}). Additionally, thanks to the rich dependency model of the SCG, we were able to easily adjust the similarity search algorithm and discard results found within same parent in order to focus on less obvious cases. Such analysis would not be feasible based on CCN since methods are not represented in the CCN model, limiting similarity computations to the class level only. Similarly, the CG model lacks parent-child relations like the SCG \textit{DECLARATION} available in the SCG model. 
It is worth to mention that more complex algorithms for finding code similarities with the SCG model can be implemented in a~similar fashion, however, it is out of scope of this paper.

\subsection{Answering research questions}
\label{sec:answering-research-questions}

\textbf{Answer to RQ1: Does the SCG model enhance software comprehension capabilities in comparison with the Class Collaboration Network and the Call Graph models?}

In the case of project structure comprehension, we have combined the computation of project metrics with data analysis based on the split-apply-combine approach \cite{Wickham2011split}. The richness of the full SCG model allowed us to discover interesting properties of the analyzed software which could not have been found based on the Call Graph or Class Collaboration Network alone. For example, we have discovered excessive use of variables in the spring-boot project and localized the relevant places in the project (Section \ref{sec:project-overview}). Additionally, we have shown a power law distribution of node degrees in software projects and focused on finding outliers. Thanks to the richness of the SCG model, we could identify all unique hub outliers for both classes (e.g., \textit{ActorRef}) and methods (e.g., \textit{ActorRef\#!}).

In the case of top important code entities, we have shown that the full SCG model, Class Collaboration Network, and the Call Graph model provide different perspectives on the project structure. Each perspectives is useful and the SCG model supports all of them, being a superset of both CG and CCN.

Unique properties of the SCG model (such as the required \textit{location} property) allow interactive source browsing in the full context of detailed dependencies and code entities that are neither with standard Call Graph analysis nor Class Collaboration Network (Section \ref{sec:reachability}).  

\textbf{Answer to RQ2: Does SCG-based data analysis enable actionable software comprehension insights?}
Several project comprehension activities presented earlier led to useful insights regarding code problems and possible solutions. In section \ref{sec:project-overview} we discovered an excessive number of variables in the \textit{spring-boot} project. A~closer look into the source code allowed us to understand the issue and propose the refactoring exercise to apply \textit{final} modifier to reduce accidental mutability and in consequence improve the code readability. 

In Section \ref{sec:project-structure-comprehension}, which focuses on project structure comprehension, we identified an anomaly in the \textit{Retrofit} project. Specifically, the \textit{parseParameterAnnotation} method exhibited an unusually high outgoing degree. We conducted an analysis to understand the cause of this anomaly and proposed refactoring activities.

Finally, in Section \ref{sec:finding-similarities}, we were able to find similar methods in different places in the code, which is not only relevant for software comprehension, but also led to proposing an improvement of the source code by extracting the \textit{common} method (presented in Listing \ref{lst:common}) and removing code duplication. 

\section{Impact of results}
\label{sec:scg-implications}

We believe that the presented results of our research indicate that the Semantic Code Graph (SCG) represents a significant development in the field of code dependency analysis and software comprehension. For researchers and experts, this work carries several benefits:

\begin{enumerate}
    \item \textbf{Comprehensive code dependency analysis}: SCG offers a detailed source code information model providing insight into code structure, including code definitions, declarations, and their relationships. This offers significant depth of analysis which opens up new avenues for understanding software systems. We have demonstrated the usefulness of the SCG model in diverse software comprehension applications, including code structure visualization, identification of crucial software entities, investigation of code quality issues, and software mining using data analytics.
    
    \item \textbf{Practical utility}: The empirical evaluation of the SCG on real-world projects demonstrates its practical utility. This utility has been demonstrated by using SCG as foundation of several tools for software comprehension: the \textit{Graph Buddy} plugin for interactive code dependency visualization in popular IDEs, and the \textit{scg-cli} tool. Researchers can use the tools that we have provided to conduct similar experiments for the purpose of software comprehension or validation of their own models. This emphasis on empirical validation helps bridge the gap between theoretical work and real-world applications on software projects.
    
    \item \textbf{Actionable insights}: We have shown that the SCG capability has led to concrete useful insights, e.g., identification of excessive usage of mutable variables, identification of outliers with extreme number of dependencies, or finding similar methods and suggesting project refactoring actions. Researchers can build on this capability to develop automated code quality improvement tools. SCG's high level of detail enables deriving other graph-based code representation models, such as Class Collaboration Network (CCN) and Call Graph (CG), which are already provided by the \textit{scg-cli} tool. Our experiments have shown that these models provide additional insights into software structure and execution dependencies. 

    \item \textbf{External integrations}: Along with the SCG model we have provided its portable intermediate representation, tools and APIs that enable integrations with external tools such as Gephi, Jupyter Notebook or any tool capable of reading a protobuf format.
    
    \item \textbf{Community Contribution and Future Research}: SCG detailed description and protobuf based intermediate representation facilitates creation of new extractors for other programming languages and the \textit{scg-cli} tool helps in SCG data analysis, establishing a solid foundation for future research endeavors in code dependency analysis, software monitoring, maintenance, and tool development. 

    \item \textbf{Cost Reduction in Project Maintenance}: With its emphasis on practical utility, the proposed SCG model and tools have the potential to influence development of future tools that help reducing project maintenance costs by expediting activities related to software comprehension. 
    
\end{enumerate}

In conclusion, the work on the Semantic Code Graph offers a rich source of inspiration and practical insights for potential researchers and experts. It encourages a shift toward empirical validation, visualization, tool development, and collaborative efforts in the realm of code dependency analysis and software comprehension, ultimately contributing to more efficient and effective software maintenance practices.

\section{Conclusions}
\label{sec:conclusions}

We introduced the Semantic Code Graph (SCG), a comprehensive source code information model. The SCG is a form of software network which focuses on highly detailed source code-level dependencies, capable of representing all code definitions, declarations, and the relationships between them. SCG introduces diverse properties crucial for subsequent analysis, as well as a simple yet effective intermediate representation in the form of a protobuf schema. We have described the implementation of the SCG model and the extraction of SCG data for Scala and Java languages.

The SCG model properties enable advanced analysis and visualization of source code structure, as well as integration with external tools. Practical capabilities of SCG have been proven in empirical evaluation on eleven popular open source projects written in Java and Scala. For each project we have extracted their SCG, Class Collaboration Network, and Call Graph models, and conducted diverse comparative software comprehension experiments. Thanks to the level of detail in SCG, even a simple distribution analysis of SCG node types can provide valuable insights, such as identifying excessive usage of mutable variables and suggesting project refactoring actions. 

Class Collaboration Network (CCN) and Call Graph (CG) also provide interesting perspectives on the software. CCN focuses on the high-level code structure, while CG analyzes important dependencies from a program execution perspective. Both CCN and CG can be effectively extracted from the more general SCG model. SCG, CCN, and CG exhibit a power-law distribution in each project, indicating that the software structure displays characteristics of a scale-free network, with a significant number of outlier nodes. These outlier nodes can sometimes be identified as places for direct refactoring improvements to enhance project readability and structural quality. The most important nodes in each network may be of interest for a different reasons: in the CCN network they are crucial from a project structure perspective, while in the CG  graph the represent code entities having a major impact on program execution. The SCG combines these perspectives to identify a unique set of nodes critical for overall project comprehension.

The SCG model has enabled effective interactive code dependency visualization, proving its usefulness in the \textit{Graph Buddy} plugin for popular IDEs. We have also demonstrated the utility of SCG in general software mining (discovery of similar code snippets) leveraging data analysis in external tools, such as Jupyter Notebook with Python data frames environment.

We argue that our research, which focuses on empirical experiments on real projects, helps address the gap between theoretical code representation models and their applications, as well as the surprising lack of concrete and widely used software comprehension tools in the common developer workflow. To encourage future research in this direction, i.e., one that emphasizes practical usability aspects of the underlying theoretical work, we have published our tools and focused on their usability, notably \textit{scg-cli}, a tool for SCG extraction\footnote{Currently \textit{scg-cli} supports SCG extraction for Java projects. For Scala, which requires a different approach, we have developed a compiler plugin for this purpose available at \url{https://github.com/VirtusLab/scg-scala}} and analysis. We hope that our work will contribute to the emergence of practical research and applications in the code structure analysis and software comprehension field.

\section*{Acknowledgments}
Krzysztof Borowski was supported in part by AGH University of Krakow and the "Doktorat Wdrozeniowy" program of the Polish Ministry of Science and Higher Education, and by the VirtusLab company.


\bibliographystyle{plain}
\bibliography{bibliography}

\begin{thebibliography}{10}

\bibitem{acharya2011practical}
Mithun Acharya and Brian Robinson.
\newblock Practical change impact analysis based on static program slicing for industrial software systems.
\newblock In {\em Proceedings of the 33rd international conference on software engineering}, pages 746--755, 2011.

\bibitem{Al-Saiyd2017}
Nedhal~A. Al-Saiyd.
\newblock Source code comprehension analysis in software maintenance.
\newblock In {\em 2017 2nd International Conference on Computer and Communication Systems (ICCCS)}, pages 1--5, July 2017.

\bibitem{Alanzi2021}
Rakan Alanazi, Gharib Gharibi, and Yugyung Lee.
\newblock Facilitating program comprehension with call graph multilevel hierarchical abstractions.
\newblock {\em Journal of Systems and Software}, 176:110945, 2021.

\bibitem{allen1970control}
Frances~E Allen.
\newblock Control flow analysis.
\newblock {\em ACM Sigplan Notices}, 5(7):1--19, 1970.

\bibitem{Arora2019}
Ritu Arora and Sanjay Goel.
\newblock Javarelationshipgraphs (jrg): Transforming java projects into graphs using neo4j graph databases.
\newblock In {\em Proceedings of the 2nd International Conference on Software Engineering and Information Management}, ICSIM 2019, page 80–84, New York, NY, USA, 2019. Association for Computing Machinery.

\bibitem{Arora2016}
Ritu Arora, Sanjay Goel, and R.~K. Mittal.
\newblock Using dependency graphs to support collaboration over github: The neo4j graph database approach.
\newblock In {\em 2016 Ninth International Conference on Contemporary Computing (IC3)}, pages 1--7, Aug 2016.

\bibitem{Arora2012}
Vinay Arora, Rajesh~Kumar Bhatia, and Maninder~Pal Singh.
\newblock Evaluation of flow graph and dependence graphs for program representation.
\newblock {\em International Journal of Computer Applications}, 56:18--23, 2012.

\bibitem{Backes2017}
Michael Backes, Konrad Rieck, Malte Skoruppa, Ben Stock, and Fabian Yamaguchi.
\newblock Efficient and flexible discovery of php application vulnerabilities.
\newblock In {\em 2017 IEEE European Symposium on Security and Privacy (EuroS\&P)}, pages 334--349, April 2017.

\bibitem{Bandi2013}
A.~{Bandi}, B.~J. {Williams}, and E.~B. {Allen}.
\newblock Empirical evidence of code decay: A systematic mapping study.
\newblock In {\em 2013 20th Working Conference on Reverse Engineering (WCRE)}, pages 341--350, 2013.

\bibitem{Bavota2013}
Gabriele Bavota, Andrea De~Lucia, Andrian Marcus, and Rocco Oliveto.
\newblock Using structural and semantic measures to improve software modularization.
\newblock {\em Empirical Software Engineering}, 18(5):901--932, 2013.

\bibitem{Bhattacharya2012}
Pamela Bhattacharya, Marios Iliofotou, Iulian Neamtiu, and Michalis Faloutsos.
\newblock Graph-based analysis and prediction for software evolution.
\newblock In {\em 2012 34th International Conference on Software Engineering (ICSE)}, pages 419--429, June 2012.

\bibitem{Bonacich1987}
Phillip Bonacich.
\newblock Power and centrality: A family of measures.
\newblock {\em American journal of sociology}, 92(5):1170--1182, 1987.

\bibitem{Borowski2022}
Krzysztof Borowski, Bartosz Bali{\'s}, and Tomasz Orzechowski.
\newblock Graph buddy --- an interactive code dependency browsing and visualization tool.
\newblock In {\em 2022 Working Conference on Software Visualization (VISSOFT)}, 2022.

\bibitem{Brandes2004}
Ulrik Brandes.
\newblock A faster algorithm for betweenness centrality.
\newblock {\em The Journal of Mathematical Sociology}, 25, 03 2004.

\bibitem{Brin1998}
Sergey Brin and Lawrence Page.
\newblock The anatomy of a large-scale hypertextual web search engine.
\newblock {\em Computer Networks and ISDN Systems}, 30(1):107--117, 1998.

\bibitem{Brito2020}
Aline Brito, Andre Hora, and Marco~Tulio Valente.
\newblock Refactoring graphs: Assessing refactoring over time.
\newblock In {\em 2020 IEEE 27th International Conference on Software Analysis, Evolution and Reengineering (SANER)}, pages 367--377, Feb 2020.

\bibitem{cserep2020integration}
M{\'a}t{\'e} Cser{\'e}p and Anett Fekete.
\newblock Integration of incremental build systems into software comprehension tools.
\newblock In {\em ICAI}, pages 85--93, 2020.

\bibitem{Du2021}
Xin Du, Tian Wang, Weifeng Pan, Muchou Wang, Bo~Jiang, Yiming Xiang, Chunlai Chai, Jiale Wang, and Chengxiang Yuan.
\newblock Cospa: Identifying key classes in object-oriented software using preference aggregation.
\newblock {\em IEEE Access}, 9:114767--114780, 2021.

\bibitem{Bois2004}
B.~Du~Bois, S.~Demeyer, and J.~Verelst.
\newblock Refactoring - improving coupling and cohesion of existing code.
\newblock In {\em 11th Working Conference on Reverse Engineering}, pages 144--151, Nov 2004.

\bibitem{Eick2001}
S.G. Eick, T.L. Graves, A.F. Karr, J.S. Marron, and A.~Mockus.
\newblock Does code decay? assessing the evidence from change management data.
\newblock {\em IEEE Transactions on Software Engineering}, 27(1):1--12, Jan 2001.

\bibitem{Falci2017}
Daniel Falci, Orlando Gomes, and Fernando Silva~Parreiras.
\newblock Complex networks analysis for software architecture: an hibernate call graph study.
\newblock 06 2017.

\bibitem{Ferrante1987}
Jeanne Ferrante, Karl~J. Ottenstein, and Joe~D. Warren.
\newblock The program dependence graph and its use in optimization.
\newblock {\em ACM Trans. Program. Lang. Syst.}, 9(3):319–349, July 1987.

\bibitem{fowler1997refactoring}
Martin Fowler.
\newblock Refactoring: Improving the design of existing code.
\newblock In {\em 11th European Conference. Jyv{\"a}skyl{\"a}, Finland}, 1997.

\bibitem{Galindo2020data}
Carlos Galindo, Sergio P{\'e}rez, and Josep Silva.
\newblock Data dependencies in object-oriented programs.
\newblock In {\em 11th Workshop on Tools for Automatic Program Analysis}, 2020.

\bibitem{Gomez2019}
Sergi Gomez.
\newblock {\em Centrality in Networks: Finding the Most Important Nodes}, pages 401--433.
\newblock Springer International Publishing, Cham, 2019.

\bibitem{Grove1997}
David Grove, Greg DeFouw, Jeffrey Dean, and Craig Chambers.
\newblock Call graph construction in object-oriented languages.
\newblock {\em SIGPLAN Not.}, 32(10):108--124, oct 1997.

\bibitem{Grune2012-intro}
Dick Grune, Kees van Reeuwijk, Henri~E. Bal, Ceriel J.~H. Jacobs, and Koen Langendoen.
\newblock {\em Modern Compiler Design}, pages 1--51.
\newblock Springer New York, New York, NY, 2012.

\bibitem{Guo2013}
Yang Guo, Zheng-xu Zhao, and Wei Wang.
\newblock Complexity analysis of software based on function-call graph.
\newblock In Ershi Qi, Jiang Shen, and Runliang Dou, editors, {\em The 19th International Conference on Industrial Engineering and Engineering Management}, pages 269--277, Berlin, Heidelberg, 2013. Springer Berlin Heidelberg.

\bibitem{horwitz1988interprocedural}
Susan Horwitz, Thomas Reps, and David Binkley.
\newblock Interprocedural slicing using dependence graphs.
\newblock In {\em Proceedings of the ACM SIGPLAN 1988 conference on Programming Language design and Implementation}, pages 35--46, 1988.

\bibitem{Husain2019}
Hamel Husain, Ho-Hsiang Wu, Tiferet Gazit, Miltiadis Allamanis, and Marc Brockschmidt.
\newblock Codesearchnet challenge: Evaluating the state of semantic code search.
\newblock {\em arXiv preprint arXiv:1909.09436}, 2019.

\bibitem{hyland2006scale}
David Hyland-Wood, David Carrington, and Simon Kaplan.
\newblock Scale-free nature of java software package, class and method collaboration graphs.
\newblock In {\em Proceedings of the 5th International Symposium on Empirical Software Engineering}. Citeseer, 2006.

\bibitem{Isazadeh2017}
Ayaz Isazadeh, Habib Izadkhah, and Islam Elgedawy.
\newblock {\em Techniques for the Evaluation of Software Modularizations}, pages 179--216.
\newblock Springer International Publishing, Cham, 2017.

\bibitem{Jenkins2007}
S.~Jenkins and S.R. Kirk.
\newblock Software architecture graphs as complex networks: A novel partitioning scheme to measure stability and evolution.
\newblock {\em Information Sciences}, 177(12):2587--2601, 2007.

\bibitem{Kalyur2016}
Sesha Kalyur and G.S. Nagaraja.
\newblock Paracite: Auto-parallelization of a sequential program using the program dependence graph.
\newblock In {\em 2016 International Conference on Computation System and Information Technology for Sustainable Solutions (CSITSS)}, pages 7--12, Oct 2016.

\bibitem{kan2003metrics}
Stephen~H Kan.
\newblock {\em Metrics and models in software quality engineering}.
\newblock Addison-Wesley Professional, 2003.

\bibitem{karuthedath2020}
Abdul~vahab karuthedath, Sreekutty Vijayan, and Vipin~Kumar K.~S.
\newblock System dependence graph based test case generation for object oriented programs.
\newblock In {\em 2020 International Conference on Power, Instrumentation, Control and Computing (PICC)}, pages 1--6, Dec 2020.

\bibitem{Katz1953}
Leo Katz.
\newblock A new status index derived from sociometric analysis.
\newblock {\em Psychometrika}, 18(1):39--43, 1953.

\bibitem{Kim2016}
Jinhyun Kim, HyukGeun Choi, Hansang Yun, and Byung-Ro Moon.
\newblock Measuring source code similarity by finding similar subgraph with an incremental genetic algorithm.
\newblock In {\em Proceedings of the Genetic and Evolutionary Computation Conference 2016}, GECCO '16, pages 925--932, New York, NY, USA, 2016. Association for Computing Machinery.

\bibitem{Kinable2011}
Joris Kinable and Orestis Kostakis.
\newblock Malware classification based on call graph clustering.
\newblock {\em Journal in Computer Virology}, 7(4):233--245, 2011.

\bibitem{Krinke2001}
J.~Krinke.
\newblock Identifying similar code with program dependence graphs.
\newblock In {\em Proceedings Eighth Working Conference on Reverse Engineering}, pages 301--309, Oct 2001.

\bibitem{lam2011soot}
Patrick Lam, Eric Bodden, Ondrej Lhot{\'a}k, and Laurie Hendren.
\newblock The soot framework for java program analysis: a retrospective.
\newblock In {\em Cetus Users and Compiler Infastructure Workshop (CETUS 2011)}, volume~15, 2011.

\bibitem{LaToza2010}
Thomas~D. LaToza and Brad~A. Myers.
\newblock Developers ask reachability questions.
\newblock In {\em 2010 ACM/IEEE 32nd International Conference on Software Engineering}, volume~1, pages 185--194, May 2010.

\bibitem{LaToza2011}
Thomas~D. LaToza and Brad~A. Myers.
\newblock Visualizing call graphs.
\newblock In {\em 2011 IEEE Symposium on Visual Languages and Human-Centric Computing (VL/HCC)}, pages 117--124, 2011.

\bibitem{Lehnert2011}
Steffen Lehnert.
\newblock A taxonomy for software change impact analysis.
\newblock In {\em Proceedings of the 12th International Workshop on Principles of Software Evolution and the 7th Annual ERCIM Workshop on Software Evolution}, IWPSE-EVOL '11, pages 41--50, New York, NY, USA, 2011. Association for Computing Machinery.

\bibitem{Li2013}
Bixin Li, Xiaobing Sun, Hareton Leung, and Sai Zhang.
\newblock A survey of code-based change impact analysis techniques.
\newblock {\em Software Testing, Verification and Reliability}, 23(8):613--646, 2013.

\bibitem{Li2021}
Hao Li, Tian Wang, Weifeng Pan, Muchou Wang, Chunlai Chai, Pengyu Chen, Jiale Wang, and Jing Wang.
\newblock Mining key classes in java projects by examining a very small number of classes: A complex network-based approach.
\newblock {\em IEEE Access}, 9:28076--28088, 2021.

\bibitem{Liu2021visual}
Huan Liu, Yubo Tao, Wenda Huang, and Hai Lin.
\newblock Visual exploration of dependency graph in source code via embedding-based similarity.
\newblock {\em Journal of Visualization}, 24(3):565--581, 2021.

\bibitem{Lutellier2018}
Thibaud Lutellier, Devin Chollak, Joshua Garcia, Lin Tan, Derek Rayside, Nenad Medvidovi{\'c}, and Robert Kroeger.
\newblock Measuring the impact of code dependencies on software architecture recovery techniques.
\newblock {\em IEEE Transactions on Software Engineering}, 44(2):159--181, Feb 2018.

\bibitem{Mattila2016}
Anna-Liisa Mattila, Petri Ihantola, Terhi Kilamo, Antti Luoto, Mikko Nurminen, and Heli V\"{a}\"{a}t\"{a}j\"{a}.
\newblock Software visualization today: Systematic literature review.
\newblock In {\em Proceedings of the 20th International Academic Mindtrek Conference}, AcademicMindtrek '16, pages 262--271, New York, NY, USA, 2016. Association for Computing Machinery.

\bibitem{McCabe1976}
T.J. McCabe.
\newblock A complexity measure.
\newblock {\em IEEE Transactions on Software Engineering}, SE-2(4):308--320, Dec 1976.

\bibitem{McIntosh2016}
Shane McIntosh, Yasutaka Kamei, Bram Adams, and Ahmed~E. Hassan.
\newblock An empirical study of the impact of modern code review practices on software quality.
\newblock {\em Empirical Software Engineering}, 21(5):2146--2189, 2016.

\bibitem{Mehrotra2022}
Nikita Mehrotra, Navdha Agarwal, Piyush Gupta, Saket Anand, David Lo, and Rahul Purandare.
\newblock Modeling functional similarity in source code with graph-based siamese networks.
\newblock {\em IEEE Transactions on Software Engineering}, 48(10):3771--3789, Oct 2022.

\bibitem{Meng2015}
Yulong Meng, Dong Xu, Ziying Zhang, and Wencai Li.
\newblock System dependency graph construction algorithm based on equivalent substitution.
\newblock In {\em 2015 Eighth International Conference on Internet Computing for Science and Engineering (ICICSE)}, pages 106--110, 2015.

\bibitem{Meyer2015}
P.~Meyer, Harvey Siy, and Sanjukta Bhowmick.
\newblock Identifying important classes of large software systems through k-core decomposition.
\newblock {\em Advances in Complex Systems}, 17:1550004, 04 2015.

\bibitem{Middleton2022}
Justin Middleton and Kathryn~T. Stolee.
\newblock Understanding similar code through comparative comprehension.
\newblock In {\em 2022 IEEE Symposium on Visual Languages and Human-Centric Computing (VL/HCC)}, pages 1--11, Sep. 2022.

\bibitem{myers2003software}
Christopher~R Myers.
\newblock Software systems as complex networks: Structure, function, and evolvability of software collaboration graphs.
\newblock {\em Physical review E}, 68(4):046116, 2003.

\bibitem{Nair2020}
Aravind Nair, Avijit Roy, and Karl Meinke.
\newblock Funcgnn: A graph neural network approach to program similarity.
\newblock In {\em Proceedings of the 14th ACM / IEEE International Symposium on Empirical Software Engineering and Measurement (ESEM)}, ESEM '20, New York, NY, USA, 2020. Association for Computing Machinery.

\bibitem{Newman2018networks}
M.~Newman.
\newblock {\em Networks}.
\newblock OUP Oxford, 2018.

\bibitem{Odersky2004}
Martin Odersky and al.
\newblock An overview of the scala programming language.
\newblock Technical Report IC/2004/64, EPFL Lausanne, Switzerland, 2004.

\bibitem{Odersky2016-tasty-reference}
Martin Odersky, Eugene Burmako, and Dmytro Petrashko.
\newblock Tasty reference manual.
\newblock page~21, 2016.

\bibitem{Oliveto2011}
Rocco Oliveto, Malcom Gethers, Gabriele Bavota, Denys Poshyvanyk, and Andrea De~Lucia.
\newblock Identifying method friendships to remove the feature envy bad smell: Nier track.
\newblock In {\em 2011 33rd International Conference on Software Engineering (ICSE)}, pages 820--823, May 2011.

\bibitem{Pan2018}
Weifeng Pan, Beibei Song, Kangshun Li, and Kejun Zhang.
\newblock Identifying key classes in object-oriented software using generalized k-core decomposition.
\newblock {\em Future Generation Computer Systems}, 81:188--202, 2018.

\bibitem{Porkolab2018}
Zolt\'{a}n Porkol\'{a}b, Tibor Brunner, D\'{a}niel Krupp, and M\'{a}rton Csord\'{a}s.
\newblock Codecompass: An open software comprehension framework for industrial usage.
\newblock In {\em Proceedings of the 26th Conference on Program Comprehension}, ICPC '18, pages 361--369, New York, NY, USA, 2018. Association for Computing Machinery.

\bibitem{Porkolab2018codecompass}
Zolt{\'a}n Porkol{\'a}b, Tibor Brunner, D{\'a}niel Krupp, and M{\'a}rton Csord{\'a}s.
\newblock Codecompass: an open software comprehension framework for industrial usage.
\newblock In {\em Proceedings of the 26th Conference on Program Comprehension}, pages 361--369, 2018.

\bibitem{Pourasghar2021}
Babak Pourasghar, Habib Izadkhah, Ayaz Isazadeh, and Shahriar Lotfi.
\newblock A graph-based clustering algorithm for software systems modularization.
\newblock {\em Information and Software Technology}, 133:106469, 2021.

\bibitem{Prause2008}
Christian~R. Prause and Stefan Apelt.
\newblock An approach for continuous inspection of source code.
\newblock In {\em Proceedings of the 6th International Workshop on Software Quality}, WoSQ '08, pages 17--22, New York, NY, USA, 2008. Association for Computing Machinery.

\bibitem{Riaz2009}
Mehwish Riaz, Muhammad Sulayman, and Husnain Naqvi.
\newblock Architectural decay during continuous software evolution and impact of `design for change' on software architecture.
\newblock In Dominik Slezak, Tai-hoon Kim, Akingbehin Kiumi, Tao Jiang, June Verner, and Silvia Abrah{\~a}o, editors, {\em Advances in Software Engineering}, pages 119--126, Berlin, Heidelberg, 2009. Springer Berlin Heidelberg.

\bibitem{Rochat2009}
Yannick Rochat.
\newblock Closeness centrality extended to unconnected graphs: The harmonic centrality index.
\newblock Technical report, 2009.

\bibitem{Prieto2020}
Oscar Rodriguez-Prieto, Alan Mycroft, and Francisco Ortin.
\newblock An efficient and scalable platform for java source code analysis using overlaid graph representations.
\newblock {\em IEEE Access}, 8:72239--72260, 2020.

\bibitem{Rosenberg1995}
Linda~H. Rosenberg and Lawrence~E. Hyatt.
\newblock Software quality metrics for object-oriented system environments.
\newblock 1995.

\bibitem{roy2007survey}
Chanchal~Kumar Roy and James~R Cordy.
\newblock A survey on software clone detection research.
\newblock {\em Queen's School of Computing TR}, 541(115):64--68, 2007.

\bibitem{Ryder1979}
B.G. Ryder.
\newblock Constructing the call graph of a program.
\newblock {\em IEEE Transactions on Software Engineering}, SE-5(3):216--226, May 1979.

\bibitem{savic2012community}
Milo{\v{s}} Savi{\'c}, Milo{\v{s}} Radovanovi{\'c}, and Mirjana Ivanovi{\'c}.
\newblock Community detection and analysis of community evolution in apache ant class collaboration networks.
\newblock In {\em Proceedings of the Fifth Balkan Conference in Informatics}, pages 229--234, 2012.

\bibitem{SAVIC2014}
Milo{\v s} Savi{\'c}, Gordana Raki{\'c}, Zoran Budimac, and Mirjana Ivanovi{\'c}.
\newblock A language-independent approach to the extraction of dependencies between source code entities.
\newblock {\em Information and Software Technology}, 56(10):1268--1288, 2014.

\bibitem{Savidis2022}
Anthony Savidis. and Crystallia Savaki.
\newblock Software architecture mining from source code with dependency graph clustering and visualization.
\newblock In {\em Proceedings of the 17th International Joint Conference on Computer Vision, Imaging and Computer Graphics Theory and Applications (VISIGRAPP 2022) - IVAPP}, pages 179--186. INSTICC, SciTePress, 2022.

\bibitem{Shah2016}
Michael~D. Shah and Samuel~Z. Guyer.
\newblock An interactive microarray call-graph visualization.
\newblock In {\em 2016 IEEE Working Conference on Software Visualization (VISSOFT)}, pages 86--90, Oct 2016.

\bibitem{Shu_2013}
Gang Shu, Boya Sun, Tim~A.D. Henderson, and Andy Podgurski.
\newblock {JavaPDG}: A new platform for program dependence analysis.
\newblock In {\em 2013 {IEEE} Sixth International Conference on Software Testing, Verification and Validation}. {IEEE}, mar 2013.

\bibitem{Siegmund2016}
Janet Siegmund.
\newblock Program comprehension: Past, present, and future.
\newblock In {\em 2016 IEEE 23rd International Conference on Software Analysis, Evolution, and Reengineering (SANER)}, volume~5, pages 13--20, March 2016.

\bibitem{Smit2011}
Michael Smit, Barry Gergel, H.~James Hoover, and Eleni Stroulia.
\newblock Code convention adherence in evolving software.
\newblock In {\em 2011 27th IEEE International Conference on Software Maintenance (ICSM)}, pages 504--507, Sep. 2011.

\bibitem{Srinuvasu2016}
Muttipati Srinuvasu and Poosapati Padmaja.
\newblock Unfolding and boosting based graph partitioning approach for object-oriented system.
\newblock {\em IARJSET}, 3:122--128, 07 2016.

\bibitem{Stapleton2020}
Sean Stapleton, Yashmeet Gambhir, Alexander LeClair, Zachary Eberhart, Westley Weimer, Kevin Leach, and Yu~Huang.
\newblock A human study of comprehension and code summarization.
\newblock In {\em Proceedings of the 28th International Conference on Program Comprehension}, ICPC '20, pages 2--13, New York, NY, USA, 2020. Association for Computing Machinery.

\bibitem{Suneja2020LearningTM}
Sahil Suneja, Yunhui Zheng, Yufan Zhuang, Jim Laredo, and Alessandro Morari.
\newblock Learning to map source code to software vulnerability using code-as-a-graph.
\newblock {\em ArXiv}, abs/2006.08614, 2020.

\bibitem{tip1994survey}
Frank Tip.
\newblock {\em A survey of program slicing techniques}.
\newblock Centrum voor Wiskunde en Informatica Amsterdam, 1994.

\bibitem{Tyagi2022}
Divya Tyagi, Ritu Arora, and Yashvardhan Sharma.
\newblock Application of java relationship graphs to academics for detection of plagiarism in java projects.
\newblock In Milan Tuba, Shyam Akashe, and Amit Joshi, editors, {\em ICT Systems and Sustainability}, pages 761--772, Singapore, 2022. Springer Nature Singapore.

\bibitem{URMA2015}
Raoul-Gabriel Urma and Alan Mycroft.
\newblock Source-code queries with graph databases---with application to programming language usage and evolution.
\newblock {\em Science of Computer Programming}, 97:127--134, 2015.

\bibitem{vallee2010soot}
Raja Vall{\'e}e-Rai, Phong Co, Etienne Gagnon, Laurie Hendren, Patrick Lam, and Vijay Sundaresan.
\newblock Soot: A java bytecode optimization framework.
\newblock In {\em CASCON First Decade High Impact Papers}, pages 214--224. 2010.

\bibitem{vaucher2009tracking}
Stephane Vaucher, Foutse Khomh, Naouel Moha, and Yann-Ga{\"e}l Gu{\'e}h{\'e}neuc.
\newblock Tracking design smells: Lessons from a study of god classes.
\newblock In {\em 2009 16th working conference on reverse engineering}, pages 145--154. IEEE, 2009.

\bibitem{Walkinshaw2003}
N.~Walkinshaw, M.~Roper, and M.~Wood.
\newblock The java system dependence graph.
\newblock In {\em Proceedings Third IEEE International Workshop on Source Code Analysis and Manipulation}, pages 55--64, Sep. 2003.

\bibitem{Wheeldon2003}
R.~Wheeldon and S.~Counsell.
\newblock Power law distributions in class relationships.
\newblock In {\em Proceedings Third IEEE International Workshop on Source Code Analysis and Manipulation}, pages 45--54, Sep. 2003.

\bibitem{Wickham2011split}
Hadley Wickham.
\newblock The split-apply-combine strategy for data analysis.
\newblock {\em Journal of statistical software}, 40:1--29, 2011.

\bibitem{Xia2018}
Xin Xia, Lingfeng Bao, David Lo, Zhenchang Xing, Ahmed~E. Hassan, and Shanping Li.
\newblock Measuring program comprehension: A large-scale field study with professionals.
\newblock {\em IEEE Transactions on Software Engineering}, 44(10):951--976, Oct 2018.

\bibitem{Yamaguchi2014}
Fabian Yamaguchi, Nico Golde, Daniel Arp, and Konrad Rieck.
\newblock Modeling and discovering vulnerabilities with code property graphs.
\newblock In {\em 2014 IEEE Symposium on Security and Privacy}, pages 590--604, May 2014.

\bibitem{Yousfi2014}
Siham Yousfi and Dalila Chiadmi.
\newblock Graph file format for etl.
\newblock In {\em 2014 Second World Conference on Complex Systems (WCCS)}, pages 189--195, Nov 2014.

\bibitem{Yu2023}
Dongjin Yu, Quanxin Yang, Xin Chen, Jie Chen, and Yihang Xu.
\newblock Graph-based code semantics learning for efficient semantic code clone detection.
\newblock {\em Information and Software Technology}, 156:107130, 2023.

\bibitem{Zhao2001}
Jianjun Zhao.
\newblock Applying program dependence analysis to java software.
\newblock 06 2001.

\end{thebibliography}

\end{document}